\pdfoutput=1
\documentclass{JINST}
\let\ifpdf\relax
\usepackage[font=footnotesize]{subfig}
\usepackage{amsmath}
\usepackage{eufrak}
\providecommand{\color}[2][1]{}

\title{Noise reduction in muon tomography for detecting high density objects 
}

\author{M.~Benettoni$^{a}$, G.~Bettella$^b$, G.~Bonomi$^c$, G.~Calvagno$^d$, P.~Calvini$^e$, P.~Checchia$^{a}$, G.~Cortelazzo$^d$, L.~Cossutta$^{ab\dagger}$, A.~Donzella$^c$, M.~Furlan$^{ab}$, F.~Gonella$^a$, M.~Pegoraro$^a$, A.~Rigoni~Garola$^{ab}$, P.~Ronchese$^{ab}$, S.~Squarcia$^e$, M.~Subieta$^c$, S.~Vanini$^{ab}$, G.~Viesti$^{ab}$, P.~Zanuttigh$^d$, A.~Zenoni$^c$ and G.~Zumerle$^{ab}$\\
\llap{$^a$}Istituto Nazionale di Fisica Nucleare, sezione di Padova, Padova, Italy\\ 
\llap{$^b$}Department of Physics and Astronomy ``G.~Galilei'', Universit\`a degli Studi di Padova, Padova, Italy\\
\llap{$^c$}Department of Mechanical and Industrial Engineering, Universit\`a di Brescia, Brescia, and INFN Sezione di Pavia, Pavia, Italy\\
\llap{$^d$}Department of Information Engineering, Universit\`a degli Studi di Padova, Padova, Italy\\
\llap{$^e$}Department of Physics, Universit\`a di Genova, and INFN Sezione di Genova, Genova, Italy \\
\llap{$\dagger$} now at Nucleco S.p.A.,
Roma, Italy\\
E-mail: \email{gianni.zumerle@unipd.it}
}

\abstract{The muon tomography technique, based on multiple Coulomb scattering of cosmic ray muons, has been proposed as a tool to detect the presence of high density objects inside closed volumes.
In this paper a new and innovative method is presented to handle the density fluctuations (noise) of reconstructed images, a well known problem of this technique.
The effectiveness of our method is evaluated using experimental data obtained with a muon tomography prototype located at the Legnaro National Laboratories (LNL) of the Istituto Nazionale di Fisica Nucleare (INFN). 
The results reported in this paper, obtained with real cosmic ray data, show that with appropriate image filtering and muon momentum classification, the muon tomography technique can detect high density materials, such as lead, albeit surrounded by light or medium density material, in short times.
A comparison with algorithms published in literature is also presented.

}
\providecommand{\muL}{\ell}

\begin{document}


\section{Introduction}
About ten years ago a novel technique to obtain tomographic images of material density of extended volumes, based on the measurements of the scattering of cosmic ray muons crossing the volume under inspection, has been proposed~\cite{borozdin}.\\
The variance of the scattering angle of a charged particle, projected on a plane containing the incoming particle direction, as a function of the radiation length $X_0$ of a material and its thickness $\muL$ can be expressed as in ~\cite{pdg}, Eq.~$30.15$. For high energy muons and low thickness this can be approximated as:
\providecommand{\B}{b}
\begin{equation}
\label{eq:sigma}
\sigma^2 \approx \frac{\B^2}{p^2}\left(\muL \lambda\right)
\end{equation}
where $\B=$13.6 MeV/c and $p$ is the particle momentum.
The quantity $\lambda=1/X_0$~~$(\mbox{rad}^{2} / \mbox{length})$ will be hereafter referred to as "scaled scattering density" (ssd) \footnote{This definition of $\lambda$ differs from the one used in Ref.~\cite{schultz} by the dimensionless factor $\left(\frac{15\mbox{(MeV/c)}}{p_0\mbox{(GeV/c)}}\right)^2$ where $p_0=3$ GeV/c.}.
As a reference in Table~\ref{table:scatt_dens} we report the computed scaled scattering density values of some elements evaluated from radiation length values reported in \cite{tsai}.

\begin{table}[h]
\begin{center}
\caption[Scaled scattering densities]{Scaled scattering density values (ssd) of some elements, ordered by increasing atomic number, evaluated from radiation length values reported in \cite{tsai}.}
\label{table:scatt_dens}
\begin{tabular}{l|c c c c c c}
\hline
                              & Al   & Fe   & Cu   & W    & Pb   & U\\
\hline
Atomic number                 & 13   & 26   & 29   & 74   & 82   & 92   \\
Mass density  (g/cm$^{3})$         & 2.7  & 7.9  & 8.9  & 19.3 & 11.3 & 19.0 \\
ssd (rad$^{2}$/cm) & 0.11 & 0.57 & 0.70 & 2.9  & 1.8  & 3.1 \\
\hline
\end{tabular}
\end{center}
\end{table}
Being the scaled scattering density proportional to the mass density and almost linearly proportional to the atomic number (Z), dense materials with high Z have the best chance to be detected using this technique.
Consequently muon tomography has been proposed~\cite{musteel} as a method to detect radioactive 
sources hidden in scrap metal containers ("orphan sources") when they are completely screened by heavy metal shields. 
In this case the radiation detector portals, commonly employed in the steel industry using scrap metal, are blind and do not trigger any alarm. 
On the contrary, a muon tomography system is able to detect the presence 
of the heavy metal shield of the radioactive source even in a dense 
metal scrap surrounding. \\
One of the greatest concerns of this technique is the presence of non negligible statistical noise in the scaled scattering density of reconstructed images. The problem is especially evident when the acquisition time is very short (\emph{i.e.} low muon statistics). This is typically the case when adapting the technique to an industrial application. Statistical noise has strong contributions from measurement errors and inefficiency of the convergence procedures. In the present paper we will address this issue.


\section{Tomographic reconstruction}
The goal of the tomographic reconstruction is to obtain a three-dimensional distribution of the scaled scattering density of the material contained in the inspected volume.
For this purpose, the space
is divided into finite volume elements called voxels, each of them assumed to have uniform scaled scattering density. 
The scaled scattering density values set $\{\lambda\}$ is estimated through the maximization of a proper likelihood function. 

\begin{figure}[htbp!]
\centering
\includegraphics[width=0.5\linewidth]{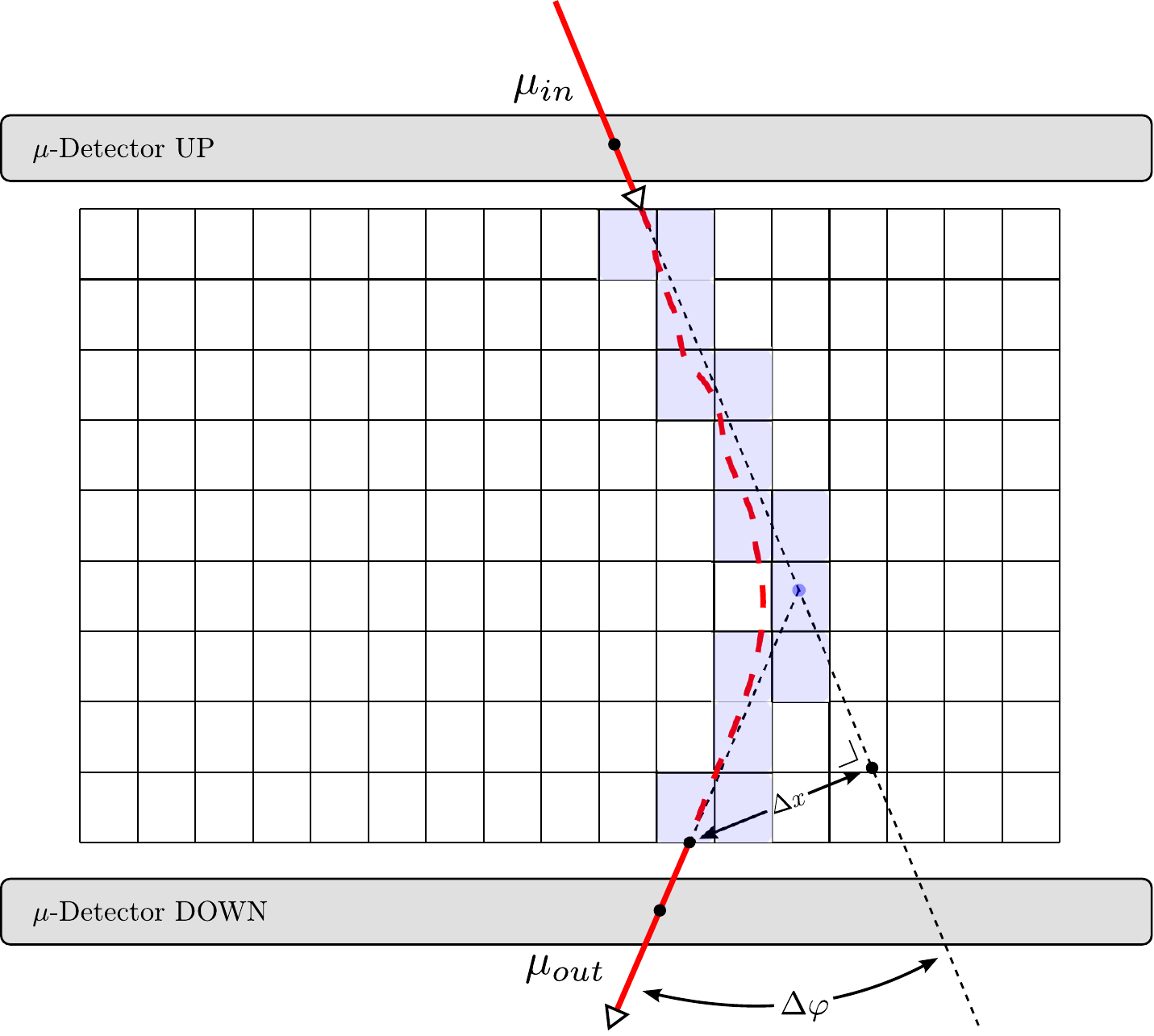}
\caption[Muon scattering sketch]
{Schematic representation of the scattering of a single muon. The quantities $\Delta \varphi$ and $\Delta x$ are the input data in the likelihood function for the tomographic reconstruction. See text for details.}
\label{fig:scat_sketch}
\end{figure}

The experimental data are the measured scattering angle and displacement of the muons. They are illustrated in Fig.~\ref{fig:scat_sketch} where a schematic representation of the scattering process of a single muon is shown. For visualization clearness the figure sketches a particular case in which both the incoming and the outgoing muon directions belong to a plane perpendicular to the detectors. The plane coincides with the figure plane. Both incoming and outgoing directions are shown, 
together with the projected scattering angle $\Delta \varphi$. The displacement $\Delta x$ is also represented. In our formalism it is defined as the projected distance between the incoming muon trajectory and the point where the outgoing trajectory exits the inspected volume. 
The displacement carries important information about the average vertical position of the scattering process, therefore it improves the reconstruction quality when its measurement is included in the likelihood function.

This section summarizes the key elements of the likelihood function maximization algorithm according to the procedure of Ref.~\cite{schultz}, where the formalism is described in detail.
We implemented it with an exact treatment of the scattering process in 3-D space for any incoming muon direction and with proper considerations of the measurement errors. We report explicitly the equations that are necessary to the comprehension and to the definition of the quantities used in the following sections. We define:
\providecommand{\Di}{\Delta_i}

\begin{equation}
\Di = \left[ {\begin{array}{*{20}{c}}
{\Delta {\varphi _i}}\\
{\Delta {x_i}}
\end{array}} \right ]
\end{equation}
as the projected scattering angle and displacement measurements of the $i$--th muon in a plane containing the incoming muon direction. Their covariance matrix is
\providecommand{\Ci}{C_i}
\providecommand{\Ni}{n_i}
\providecommand{\Tij}{t_{ij}}
\providecommand{\Lij}{\ell_{ij}}
\begin{equation}
\Ci = \left[ {\begin{array}{*{20}{c}}
{\sigma_{\Delta \varphi_i}^2} & {{\sigma_{\Delta \varphi_i , \Delta x_i}}} \\
{{\sigma _{\Delta \varphi_i , \Delta x_i}}} & {\sigma_{\Delta x_i}^2}
\end{array}} \right].
\end{equation}
If the $i$--th muon crosses $\Ni$ voxels, and the average scaled scattering density of the $j$--th voxel is $\lambda_{j}$, the covariance matrix can be expressed as 

\begin{equation}
C_i = {E_i} + \frac{\B^2}{{p_i^2}}\sum\limits_{j = 1}^{{\Ni}} {{W_{ij}}{\lambda_j}}
\end{equation}
where the parameter $\B$ has already been defined for equation~\ref{eq:sigma}, $p_i$ is the particle momentum assumed to be constant along its trajectory and $E_i$ is the contribution of the measurement errors.
For the $i$--th muon, let $\Lij$ be the length of its path inside voxel $j$, and $\Tij$ the length of its path from voxel {\it j} to the point where the outgoing trajectory exits the inspected volume.
Then the matrix $W_{ij}$ can be written as: 
\begin{equation}
W_{ij} = \left[ 
\begin{array}{*{20}{c}}
\Lij & \frac{\Lij^2}{2} + \Lij \Tij \\
\frac{\Lij^2}{2} + \Lij \Tij & \hspace{0.5cm} \frac{\Lij^2}{3} + \Lij^2 \Tij + \Lij\Tij^2
\end{array} \right].
\end{equation}
Assuming a Gaussian distribution of the measured variables, the log-likelihood of the observed scattering is 
\begin{equation}
\log \mathfrak{L}_{i} = - \frac{1}{2}\left( {\log \left| {{C_i}} \right| + \Delta _i^TC_i^{ - 1}{\Delta _i}} \right) \mbox{~+~} \mbox{K}_i
\end{equation}
where $\mbox{K}_i$ represents the terms not containing  $\{\lambda\}$.
Since muons are mutually uncorrelated, the overall log-likelihood function of a given data set is: 
\begin{equation}
\log \mathfrak{L} = - \frac{1}{2}\sum\limits_i {\left( {\log \left| {{C_i}} \right| + \Delta _i^TC_i^{ - 1}{\Delta _i}} \right)} \mbox{~+~} \sum_i \mbox{K}_i.
\end{equation}
The scaled scattering density set $\{\lambda\}$ that best reproduces the observed measurements is obtained by maximizing the likelihood function using an iterative process ~\cite{schultz}. 
For a given approximate scaled scattering density set $\{\lambda^{(n)}\}$ at iteration step $n$, a new set
$\{\lambda^{(n+1)}\}$ is obtained at step $n+1$ using the expression:

\providecommand{\Mj}{{m_j}}

\begin{equation}
\label{eq:lambda}
\lambda _j^{\left( {n + 1} \right)} = \lambda _j^{\left( n \right)} + \delta _j^{\left( n \right)} = \lambda _j^{\left( n \right)} + {\left( {\lambda _j^{\left( n \right)}} \right)^2}\frac{1}{\Mj}\sum\limits_{i,{\Lij} \ne 0} {{s_{ij}}}
\end{equation}
where $\Mj$ is the number of muons crossing the $j$--th voxel for which $\Lij$ is different from zero, and

\providecommand{\Sij}{{s_{ij}}}
\providecommand{\Sj}{{s_j}}

\begin{equation}
\Sij = \frac{1}{{2p_i^2}}\left[ {\Delta _i^TC_i^{ - 1\left( n
   \right)}{{W}_{ij}}C_i^{ - 1\left( n \right)}{\Delta _i} - Tr\left(
    {C_i^{ - 1\left( n \right)}{{W}_{ij}}} \right)} \right]
    \label{eq:sij}
\end{equation}
represents the contribution of the scattering of the $i$--th muon in the $j$--th voxel.

The plane where the scattering process is projected can be arbitrarily chosen, as long as it contains the incoming muon direction. The scattering variables projected in two orthogonal planes are uncorrelated, except for the small correlation introduced by the measurement errors on the incoming and outgoing muon directions. 
As a consequence, the four scattering variables (two angles and two displacements), measured in two orthogonal planes intersecting along the incoming muon direction, deliver the most complete information about the scattering process. For each and every muon they are the proper input of the likelihood function.


\section{Muon tomography prototype}
We collected real cosmic muon scattering data with the muon tomography prototype located at the Legnaro National Laboratories (LNL) of the Istituto Nazionale di Fisica Nucleare (INFN). We report the main characteristics of the apparatus, described in detail in Ref.~\cite{pesente}. 

\begin{figure}[htbp!]
\centerline{
\subfloat[]{\includegraphics[width=0.5\textwidth]{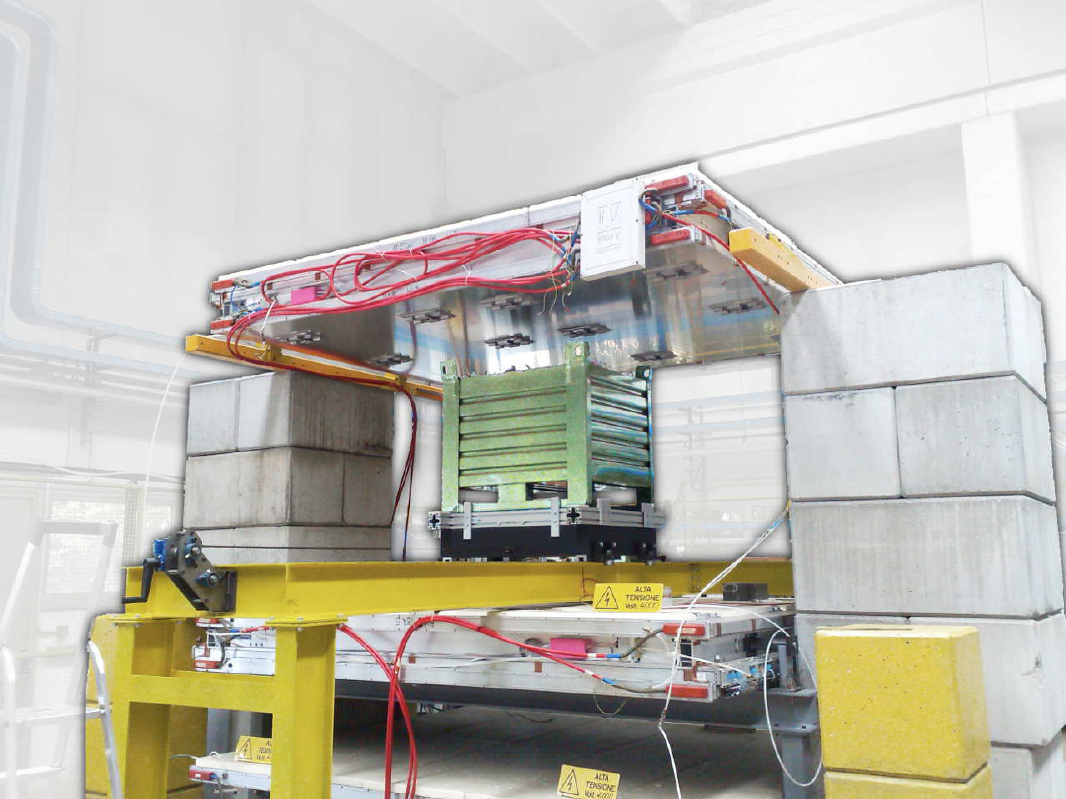} \label{fig:demonstrator_a} }
\hfill
\subfloat[]{ \includegraphics[width=0.5\textwidth]{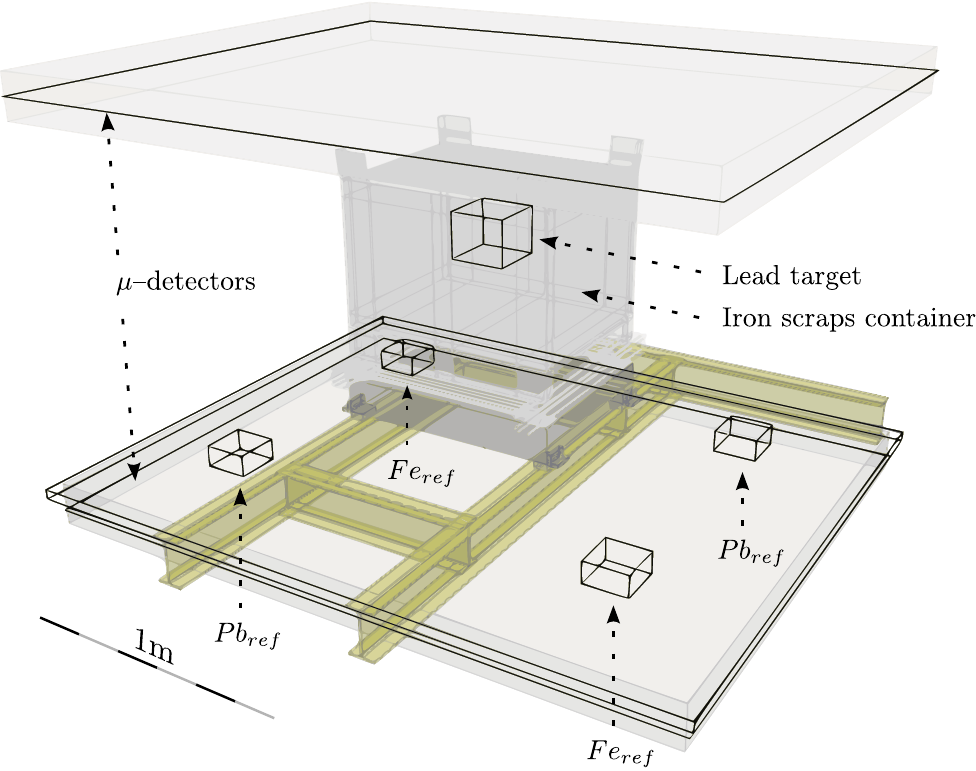} \label{fig:demonstrator_b} }
}
\caption[Prototype apparatus]
{A picture~(a) and a sketch (b) of the muon tomography prototype at the INFN LNL. Two muon detectors DT chambers of the CMS experiment (CERN) are used for tracking the muons. The large container placed on the support structure is filled with iron scrap, with a $25\times25\times20~\mbox{cm}^{3}$ lead block hidden inside; two lead and two iron reference blocks are placed near the four corners of the lower chamber.}
\label{fig:demonstrator}
\end{figure}

A photograph and a sketch of the setup are shown in Fig.~\ref{fig:demonstrator}. Two spare muon detectors built for the CMS experiment~\cite{chatrchyan} (CERN, Geneva) are used for tracking the muons. Details about the detectors can be found in~\cite{benettoni}.
The two $300\times250~\mbox{cm}^2$ detectors are placed horizontally, $160~\mbox{cm}$ apart, enclosing a volume of $11~\mbox{m}^3$.
A sturdy structure allows heavy objects to be placed inside this volume. The large container placed on the support structure is filled with iron scrap.
Two lead and two iron $20\times20\times10~\mbox{cm}^{3}$ reference blocks are placed near the four corners of the lower chamber.
A $25\times25\times20~\mbox{cm}^{3}$ lead block is hidden inside the scrap metal to test the system detection capability.

It is necessary to underline that the detectors used to assemble the prototype were designed for the needs of the CMS experiment and are not optimized for muon tomography.
They measure the muon impact position and its direction in two orthogonal planes, both orthogonal to the chamber surface, called $\varPhi$ and $\varTheta$ planes, with a high angular resolution in the $\varPhi$ plane and a lower resolution in the orthogonal $\varTheta$ plane.

In the $\varPhi$ plane the track is measured with 8 points, two groups of 4 points separated by a distance of about 24 cm. In the $\varTheta$ plane the track is measured with 4 points and a lever arm of only 3.9 cm.
The large uncertainty on the measurement of the incoming muon direction in the $\varTheta$ plane entails a large error on the displacement measurement in this plane. 
As a consequence, the projected scattering angle in this plane and the corresponding displacement information, if used in the likelihood function, do not improve the tomographic reconstruction.
So, for the present analysis, we use the scattering angle $\Delta\varphi$ and the displacement $\Delta x$ only as input data for the likelihood function, whereas the measurements in the $\varTheta$ plane are used only to track the muon. 
With this restriction the presented results are certainly slightly worse compared to results that could be obtained using optimized detectors.

\begin{figure}[htbp!]
\centering
\includegraphics[width=0.73\linewidth]{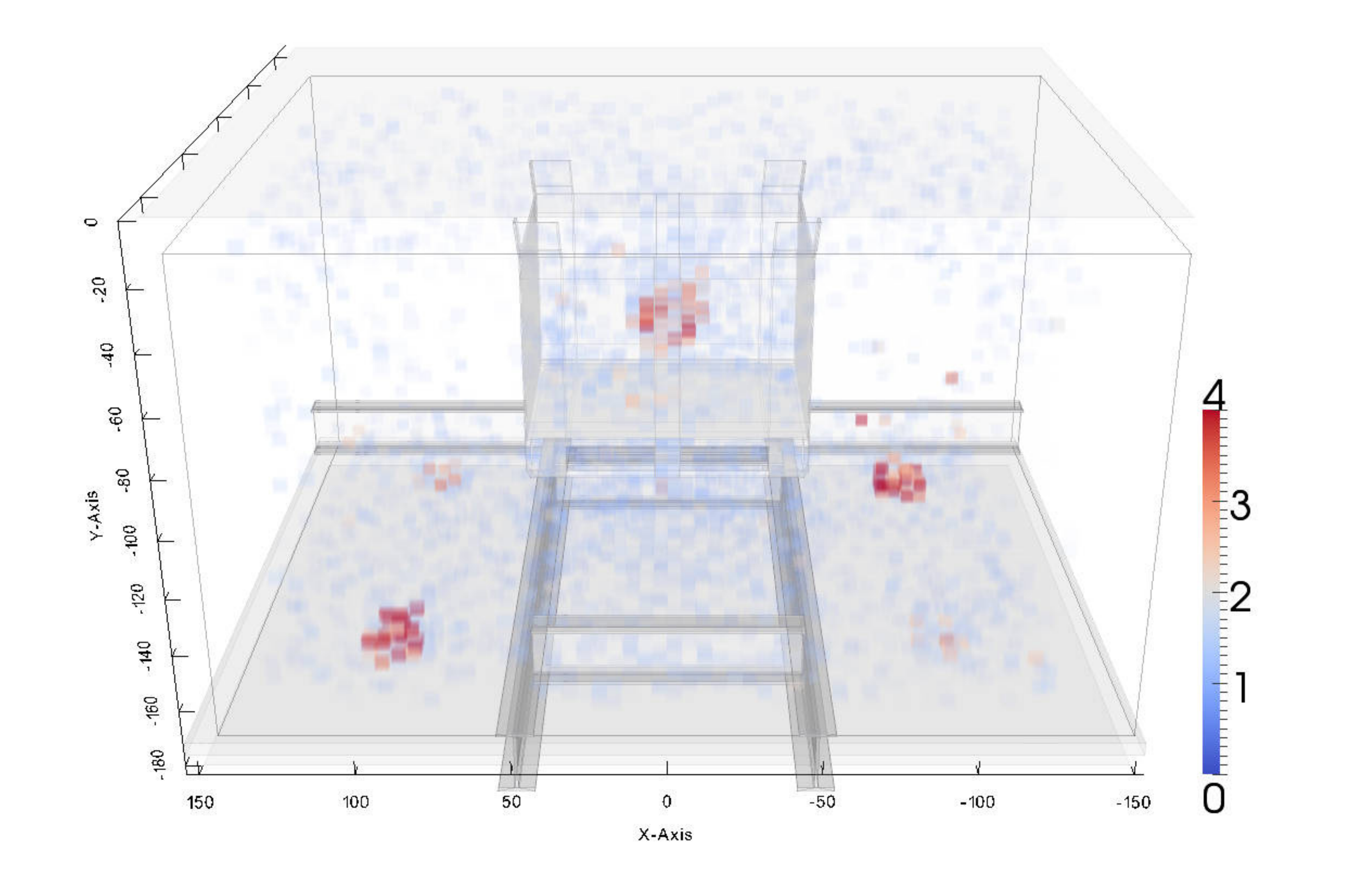}
\caption[20~min reconstruction image]
{A tomographic reconstruction of the setup shown in Fig.~\ref{fig:demonstrator} and described in the text, based on real cosmic muon data collected by the LNL prototype~(acquisition time 20~min, cubic voxels of 5 cm side). The sketch  shown in Fig.~\ref{fig:demonstrator}~(b) is faintly superimposed for clarity.}
\label{fig:reco_image}
\end{figure}

Fig.~\ref{fig:reco_image} shows a tomographic image reconstructed using the data collected by the prototype and the scattering input data described above.
An acquisition time of 20 min and a voxel size of 5 cm have been used. A sketch of the supporting structure is superimposed to the image.
The lead block hidden in the scrap metal and the two lead reference blocks can be easily recognized in the picture.

\section{Target identification algorithm}
We want to use muon tomography to assess or exclude the presence of high density objects in non accessible volumes. Our goal is to reach a high detection efficiency while keeping a low rate of false positives with short inspection time.
We studied the performance of the technique for inspection times spanning from 5~min down to 15 s using a single-blind testing procedure.

A threshold algorithm is used to spot a dense material object (the target) inside the volume under inspection.
More complex algorithms were tested without significantly better outcomes.
The target is declared to be found if at least one voxel in the volume presents a scaled scattering density value larger than a selected threshold.
Regions containing the reference blocks are excluded from the search.

This implements a binary classifier described in terms of the probability of positive/negative response given a single blind test.
We define the true positive fraction (TPF) as the probability to trigger the alarm when the target is present. This quantity is commonly called ``sensitivity''.
The probability to declare the absence of the target in samples not containing it, i.e. the true negative fraction (TNF), is usually called ``specificity'' ~\cite{ROC}. Sometimes the same information is given in terms of the false positive fraction (FPF = 1 - TNF).
To measure the true positive and negative fractions many independent data samples must be available, both containing and not containing the hidden target.
For each inspection time, two sets of samples are built.
We call ``A'' the reconstructed image set of the data samples hiding the target (a lead block), and ``B'' the image set without the target. 
We use about 400 images per set when the acquisition time is one minute or less. For longer times, a proportionally smaller number of images is used. Images are reconstructed using cubic voxels of 10 cm side.
The sensitivity (TPF) is obtained counting the positives in set A at a given threshold value. Alike, the count of the negatives in set B gives the specificity (TNF).

\begin{figure*}[htbp!]
\centerline{
\subfloat[]{ \includegraphics[width=0.5\textwidth]{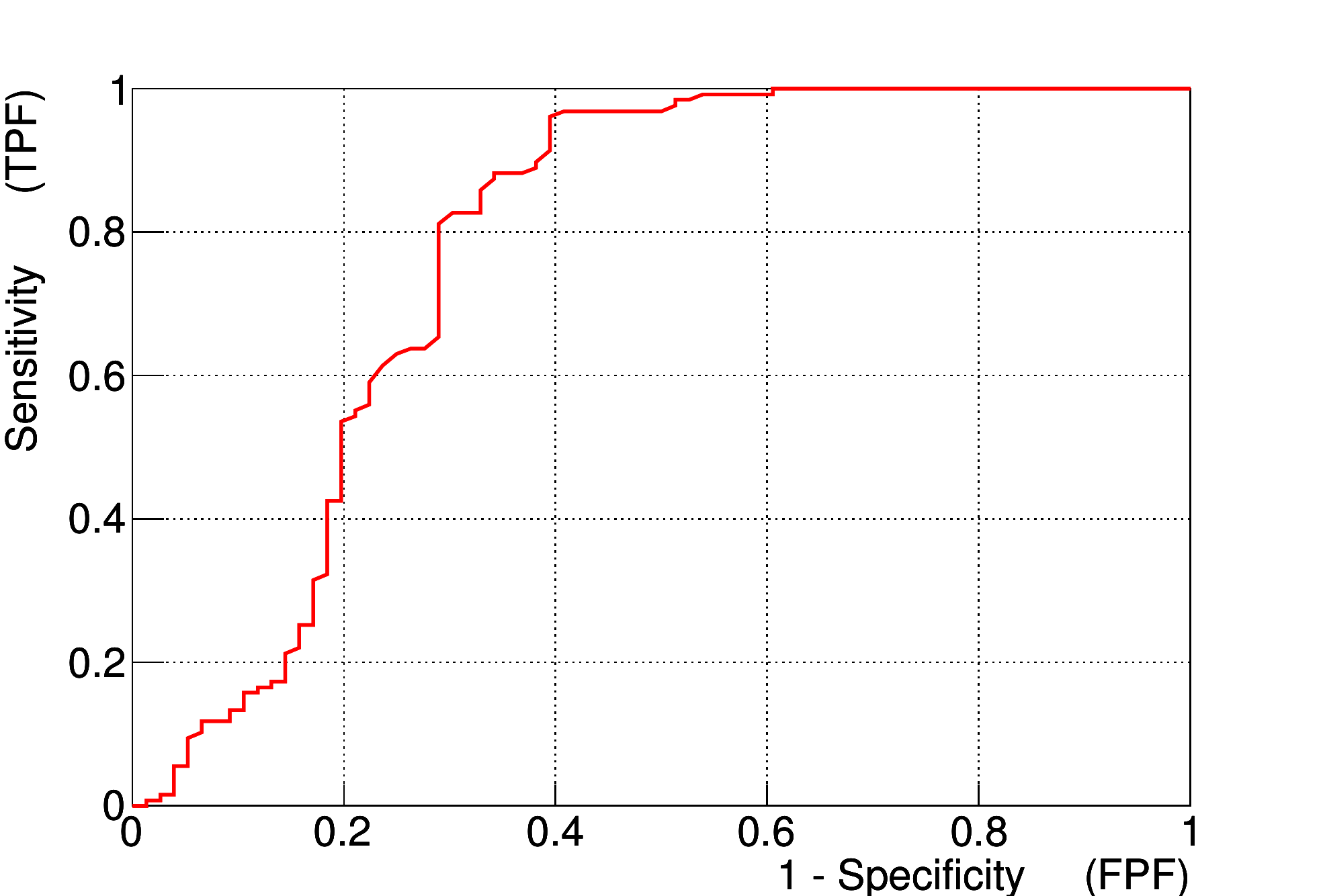} \label{fig:roc_vs_TPN_a} }
\hfill
\subfloat[]{ \includegraphics[width=0.5\textwidth]{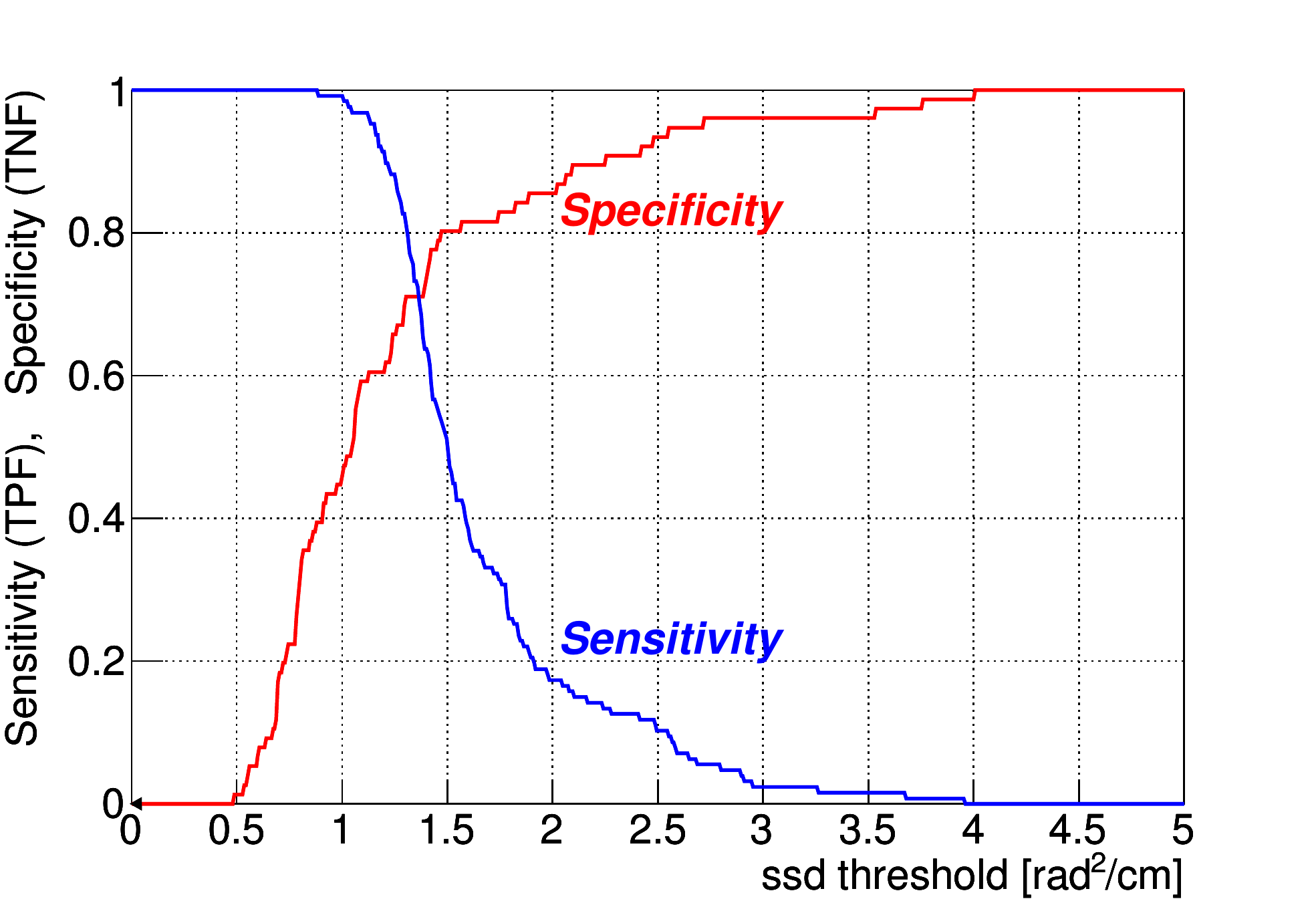} \label{fig:roc_vs_TPN_b} }
}
\caption[ROC versus TPN plots]
{(a) ROC (Receiver Operating Characteristic) and (b) TPN plot. Inspection time: 5~min. Setup of Fig.~\ref{fig:demonstrator}.
}
\label{fig:roc_vs_TPN}
\end{figure*}

The performance of a binary discrimination system is commonly represented using ROC (Receiver Operating Characteristic) curves, \textit{i.e.} the true positive fraction (sensitivity) versus the false positive fraction (1-specificity) for a range of threshold values.
An example is shown in Fig.~\ref{fig:roc_vs_TPN}~(a) where the ROC curve is obtained with samples of plain reconstructed images of 5~min acquisition time, with the setup of Fig.~\ref{fig:demonstrator}. 

There are apparent disadvantages of the ROC plot. The decision thresholds are not displayed in the plot, though they are used to generate the graph. The ROC curve collapses on the two axis unless a sufficient overlap of the two fractions, the true positive one (TPF) and the true negative (TNF) or false
positive one (FPF), exists.
For those two reasons we prefer to draw separately on the same plot the  sensitivity and specificity values as a function of the threshold. 
In the following we shall call this representation the TPN plot (True Positives and Negatives plot). An example is reported in Fig.~\ref{fig:roc_vs_TPN}~(b), obtained using the same images of Fig.~\ref{fig:roc_vs_TPN}~(a).

When sensitivity and specificity are both close to one, the system approximates a perfect predictor. We want to develop algorithms capable to reach this condition in short inspection times. 
When comparing different algorithms, the one giving a perfect prediction within the largest threshold range  will be preferred, since it is more robust against variations of the materials to inspect or other unexpected effects.
 
In order to quantify the algorithms performance, we define the ``separation ratio'' as the ratio of thresholds ($\mbox{\large \emph t\footnotesize\emph h}$) for which sensitivity (TPF) and specificity (TNF) are equal to 0.95
\begin{equation}
R=\frac{ \mbox{\large \emph t\footnotesize\emph h}( {\mbox{\footnotesize TPF=0.95}} ) }
{\mbox{\large \emph t\footnotesize\emph h}( {\mbox{\footnotesize TNF=0.95}} )} .
\end{equation}
The error on the ratio R will be obtained by estimating the binomial error, at 68\% confidence level, of the TPF and TNF curves.

In the following sections we will compare TPN plots and separation ratios to identify the best algorithm for muon tomography in terms of identification performance.

\section{Noise reduction}
The scattering angle distribution is well approximated by a Gaussian function only if the muon beam is monochromatic.
Cosmic ray muons present quite a broad momentum distribution \cite{naumov,cosmics} and the momentum of individual muons is usually unknown, because its measurement is quite difficult.
As a consequence the experimental distribution of the scattering variables is the sum of Gaussian distributions with very different width and hence largely non-Gaussian.
Nevertheless the algorithm we use for the tomographic reconstruction assumes Gaussian statistics, with variance equal to the experimental one, {\it i.e.} Eq.~(\ref{eq:sigma}) is replaced by

\begin{equation}
\sigma _{\exp }^2 = {\B^2}\left\langle {\frac{1}{{{p^2}}}} \right\rangle \left( {\ell{\lambda}} \right).
\end{equation}

The quantity $\left\langle 1/p^2\right\rangle$ can be computed from the known spectrum of cosmic muons.
It can effectively be used as a normalization parameter for the scaled scattering density. 
The use of a Gaussian approximation with  $\left\langle 1/p^2\right\rangle$ as a fixed parameter has to two important drawbacks.

\begin{itemize}

\item The reconstructed scaled scattering density is biased because the value $\left\langle 1/p^2\right\rangle$ can be modified by absorption of low energy muons~\cite{pesente}.
When crossing the denser part of the volume, muons can be stopped before reaching the lower detector. 
The average momentum value of the surviving muons is larger. 
Since the reconstruction algorithm does not consider this physical effect, it underestimates the average momentum of muons crossing the denser materials and consequently it underestimates scattering densities.

\item The large non-Gaussian tails of the scattering angle distribution generate fluctuations in the reconstructed densities.
They are especially relevant when muon statistics is poor due to short inspection time.

\end{itemize}
We address these issues in the next paragraphs.

\subsection{Image filtering}
\label{sec:noise}
\begin{figure*}[htbp!]
\centerline{
\subfloat[]{ \includegraphics[width=0.5\textwidth]{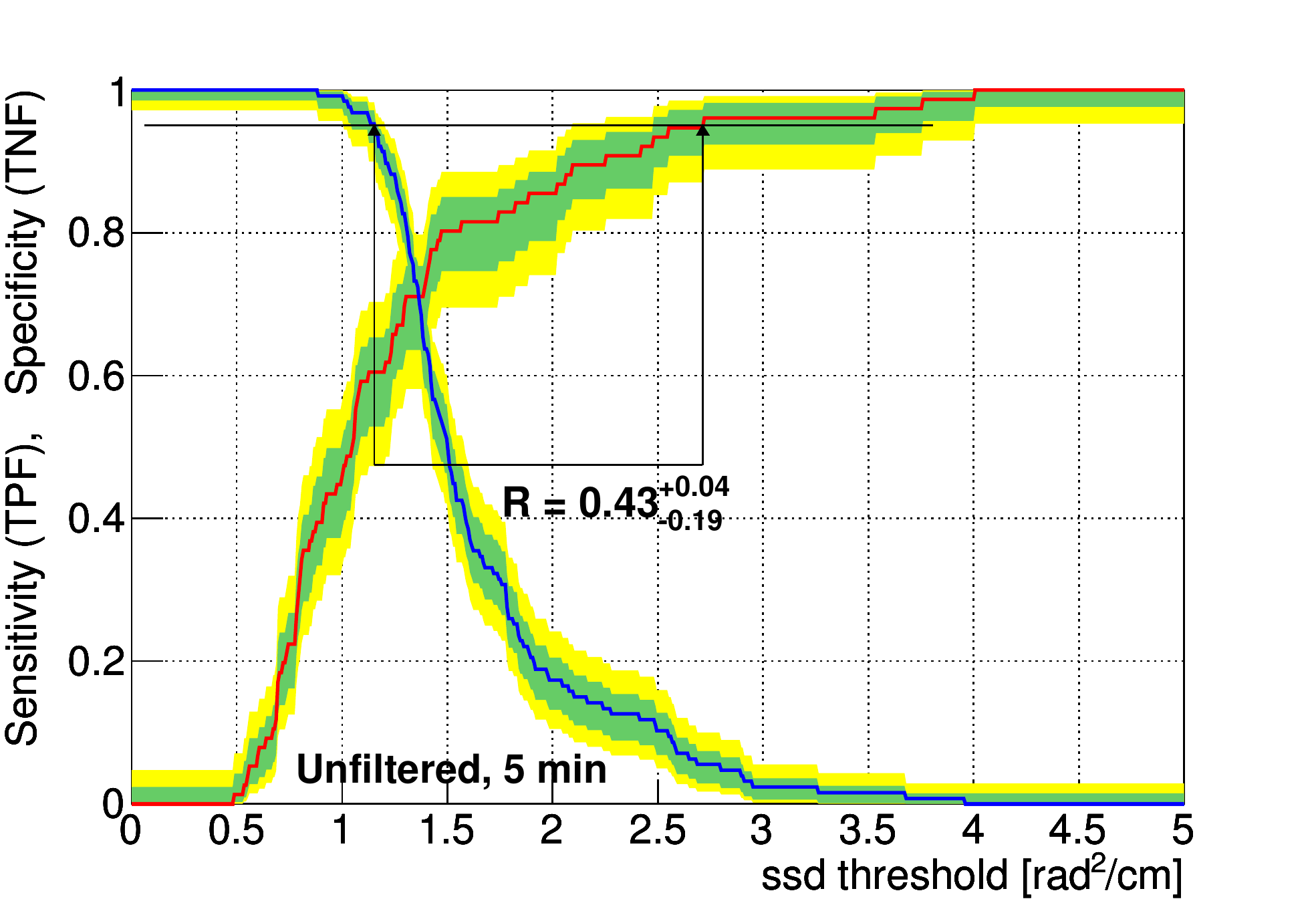} }
\hfill
\subfloat[]{ \includegraphics[width=0.5\textwidth]{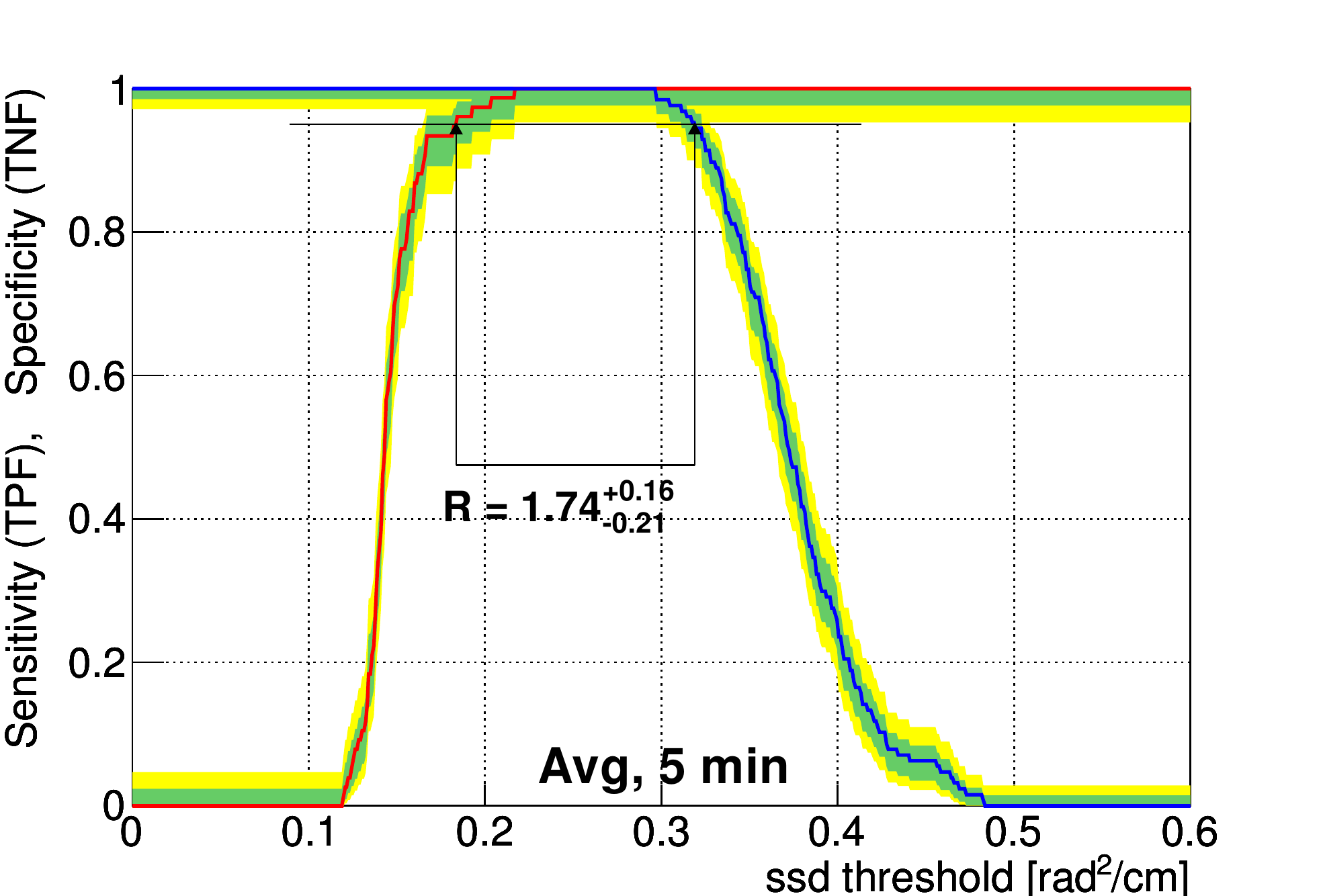} }
}
\caption[Filter effect and threshold normalization]
{(a) TPN plot for 5~min inspection time before filtering and (b) after filtering with a moving average filter, cubic kernel of $3 \times 3 \times 3$ voxels . The darker band (green) represents the binomial error at $68\%$ confidence level, the lighter band (yellow) at $95\%$ confidence level. The arrows embrace the separation range. The separation ratio is reported below the arrows.}
\label{fig:Filter_vs_NoFilter}
\end{figure*}

Noise present in digital images is usually reduced using filters. Filtered images are obviously different from the original ones, the high density values are reduced and the space information is blurred. However our goal is to discriminate high and low density objects, not to obtain an absolute measurement of the scattering density nor a determination of the object shape. Therefore the above mentioned effects are not important. What is important for our purpose is that TPN plots obtained with filtered images show a dramatic improvement compared with the unfiltered ones. 
An example is reported in Fig.~\ref{fig:Filter_vs_NoFilter} where unfiltered images are used in plot~(a) and images filtered using a moving average filter with a cubic kernel of $3 \times 3 \times 3$ voxels are used to build TPN plots in (b). The inspection time is 5~min. The darker band (green) represents the binomial error at $68\%$ confidence level, the lighter band (yellow) at $95\%$ confidence level. After filtering, the scattering density values of the region of interest are lower, but the filtered TNF and TPF curves clearly show a separation ratio greater than one. Now a prediction can be achieved for a range of thresholds. Numerically, the separation ratio moves from $R=0.43^{+0.04}_{-0.19}$ in absence of filtering to $R=1.74^{+0.16}_{-0.21}$, showing a great improvement.

The change of the scattering density values operated by the filter could make  the comparison of different TPN plots difficult. 
To overcome this problem, from now on we will normalize the threshold scale in all TPN plots, dividing it by the mean of both, $\mbox{th}( \footnotesize{\mbox{TPF=0.95}} )$ and $\mbox{th}( \footnotesize{\mbox{TNF=0.95}})$.
Obviously the R ratios are not affected by this rescaling.

\begin{figure*}[htbp!]
\centerline{
\subfloat[]{ \includegraphics[width=0.5\textwidth]{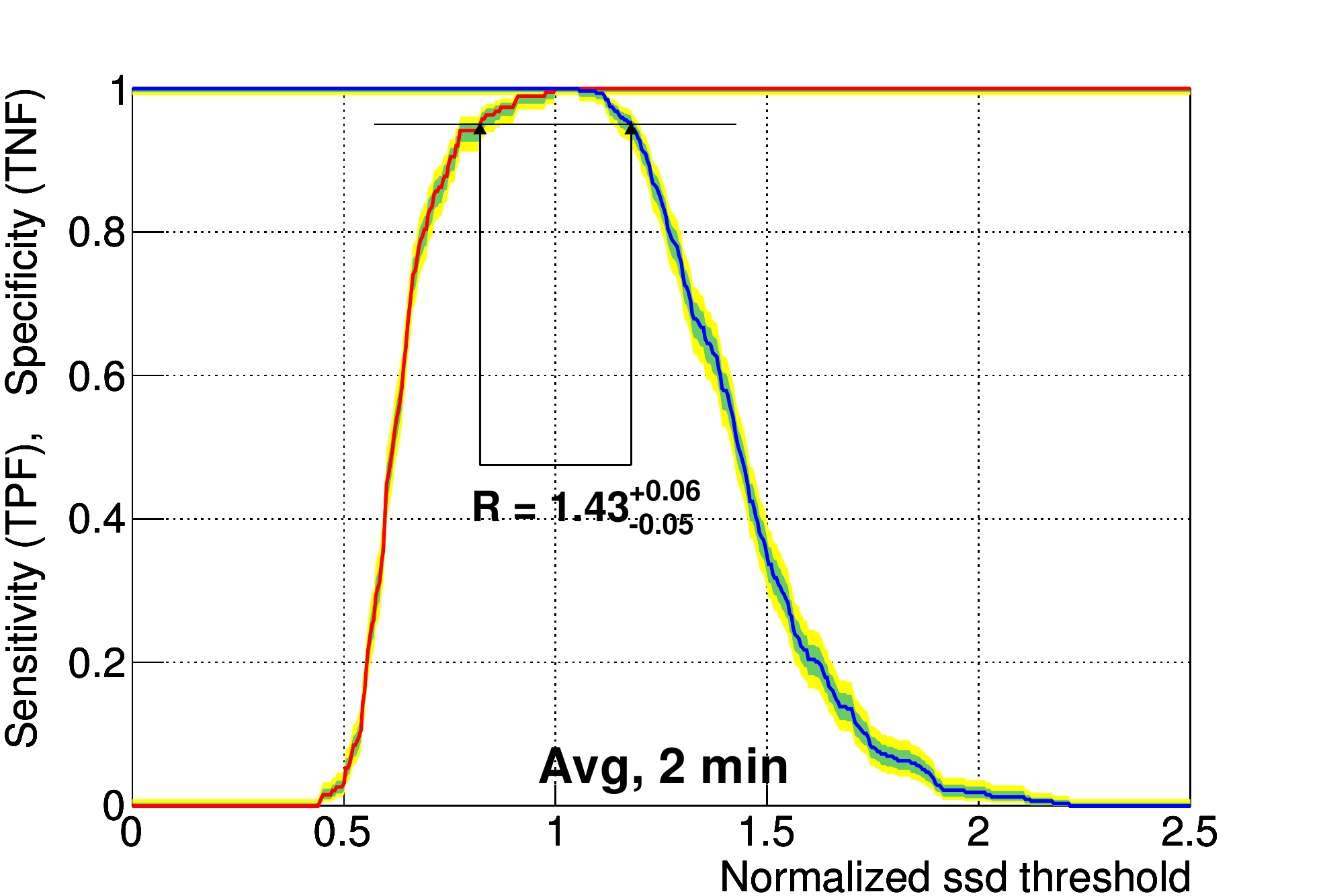} \label{fig:5-1minTPN_a} }
\hfill
\subfloat[]{ \includegraphics[width=0.5\textwidth]{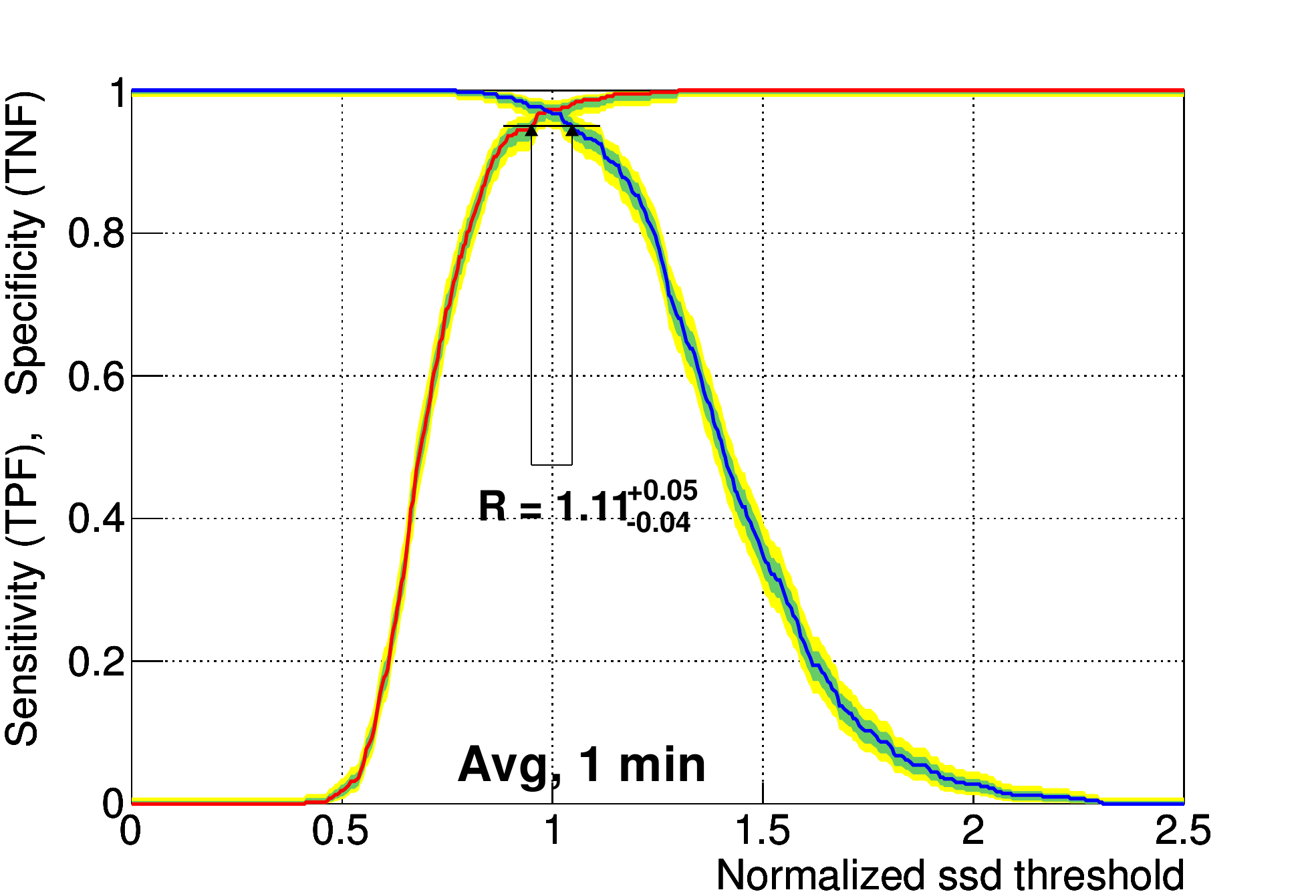} \label{fig:5-1minTPN_b} }
}
\centerline{
\subfloat[]{ \includegraphics[width=0.5\textwidth]{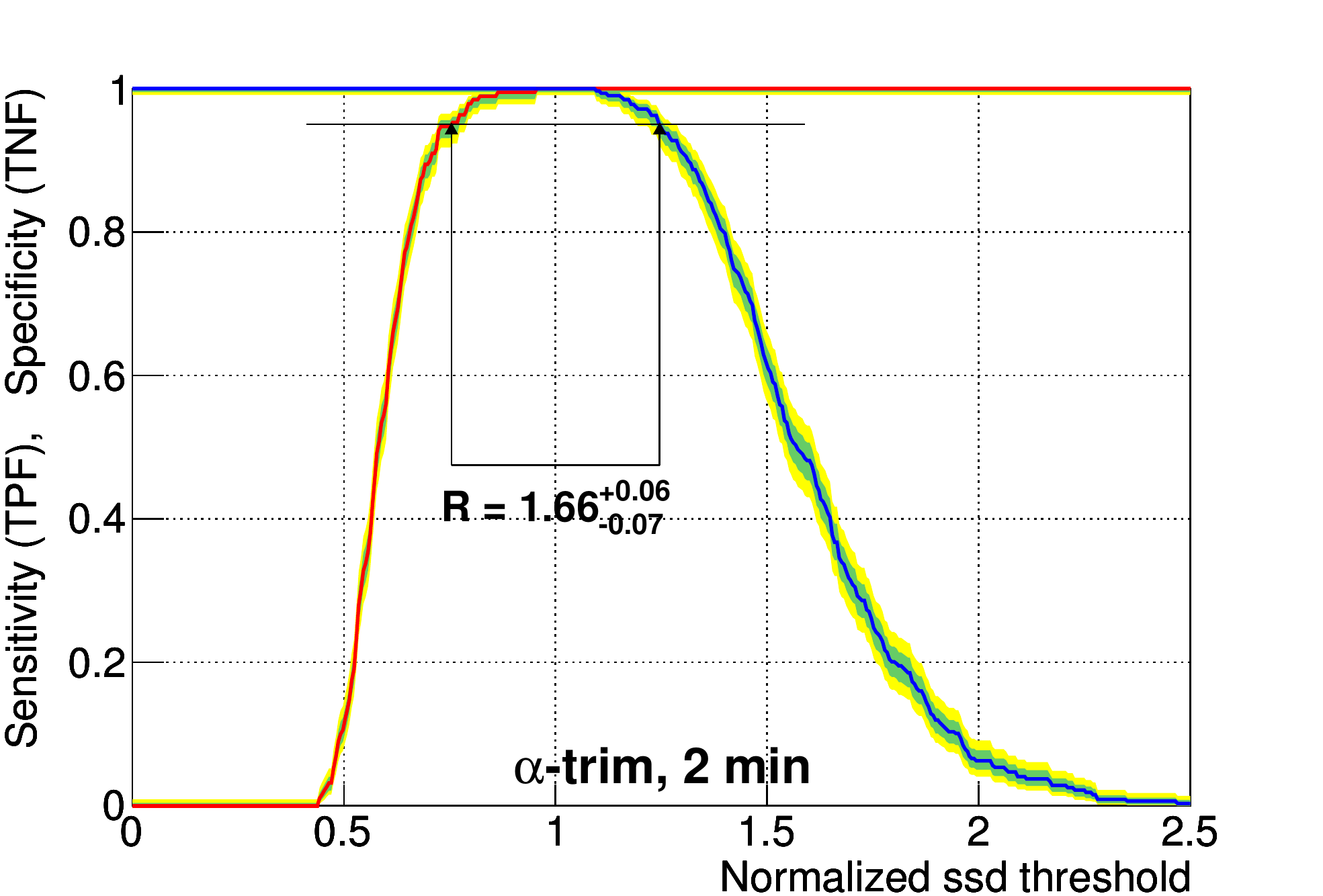} \label{fig:5-1minTPN_c} }
\hfill
\subfloat[]{ \includegraphics[width=0.5\textwidth]{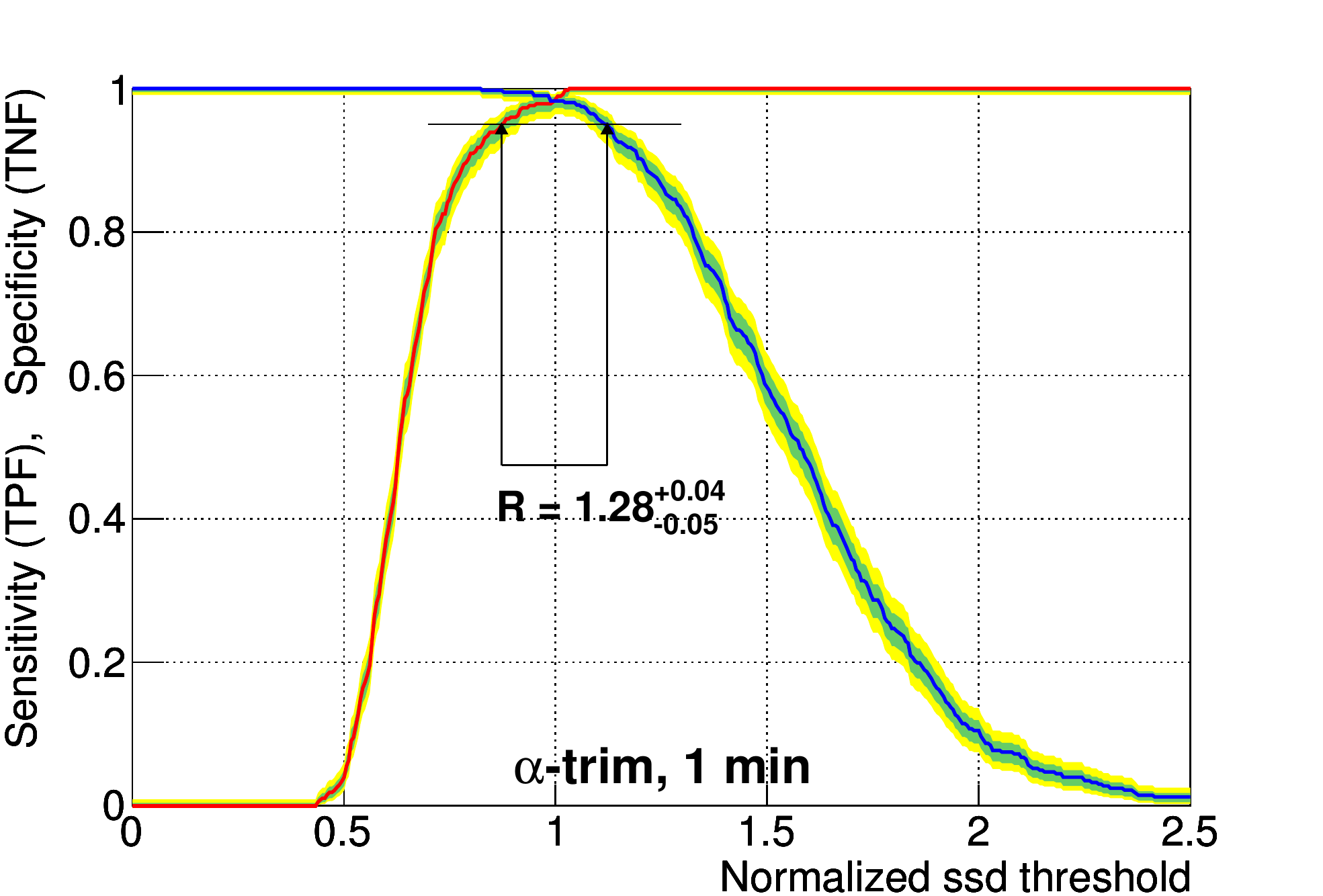} \label{fig:5-1minTPN_d} }
}
\caption[TPN curves 5-1 minutes]
{(a) TPN plot on a normalized threshold scale, using moving average filter, for a 2~min inspection time; (b) same as (a) for a 1~min inspection time. (c) TPN plot for a 2~min inspection time with $\alpha$--trim filter applied to images; (d) same as (c) for a 1~min inspection time.}
\label{fig:5-1minTPN}
\end{figure*}

We apply the moving average filter to the images obtained with a reduced inspection time. The resulting  TPN plots are shown in Fig.~\ref{fig:5-1minTPN}~(a) for 2~min inspection time and (b) for 1~min inspection time. The separation ratio is still greater then one.

More complex filters are discussed in literature. We tested the asymmetric $\alpha $--trimmed mean filter  because it is widely used for the restoration of signals and images corrupted by additive non-Gaussian noise \cite{atrim}. This filter is especially effective if the underlying noise deviates from Gaussian with impulsive components, which is the case for the scaled scattering density fluctuations of muon tomography reconstruction.
The asymmetric $\alpha$--trimmed filter is obtained by sorting voxel values inside the filter mask into ascending order, removing (trimming) a fixed fraction ($\alpha$) from the high end of the sorted set, and computing the average of the remaining values.
In our implementation ($\alpha $--trim hereafter) a weighted low-pass smoothing instead of the simple average is applied to the trimmed set.
We use $\alpha$--trim filter with a $3 \times 3 \times 3$ voxels mask, where one voxel is trimmed ($\alpha=1/27$), and a Gaussian kernel with $\sigma$ equal to the voxel size is applied. This is the default parametrization for the $\alpha$--trim filter used throughout the paper unless otherwise specified.

The $\alpha$-trim filter gives $R=1.66^{+0.06}_{-0.07}$ for a 2~min inspection time, and $R=1.28^{+0.04}_{-0.05}$ for a 1~min inspection time, as shown respectively in Fig.~\ref{fig:5-1minTPN}~(c) and (d). The result is slightly better than the one obtained with the simple moving average filter.  

\subsection{Improvements of the likelihood function}
\label{sec:Sij}
Filters are applied to images after their reconstruction.
A different and complementary strategy would be to reduce the impact of the lack of knowledge of the muon momentum during the image reconstruction process.

We propose a new algorithm to classify muon momentum based on the $\Sij$ quantities defined in \ref{eq:sij}. As equation~\ref{eq:lambda} shows, the scaled scattering density increment $\delta_j$ for the $j$--th voxel depends on the average value of the $\Sij$ quantity.
At the beginning of the image reconstruction process the densities of all voxels are set to a common value.
The same holds for the momentum of all muons.
As a consequence, at the first step of the iteration process $\Sij$ depends on the scattering variables only, its value being larger for large scattering angles. Muons with very high momentum usually present small scattering angles.
Their $\Sij$ value in equation~\ref{eq:lambda} is therefore small.
On the other hand, muons with low momentum can scatter with large angles, producing on average larger $\Sij$ values. Their contribution to $\delta_j$ can therefore be large.

We detail the algorithm in the following. Let $\left\langle \Sj \right\rangle$ be the average of $\Sij$ over the muons crossing the $j$--th voxel.
If the $i$--th muon has low momentum then $\Sij$ will tend to be larger than $\left\langle \Sj \right\rangle$ in all the voxels crossed by this muon. 
The algorithm computes the $\Sij$ and $\left\langle \Sj \right\rangle$ values and their ratios $r_{ij}=\Sij/\left\langle \Sj \right\rangle$ during the first iteration.
A muon $i$ is flagged as anomalous in voxel $j$ if $r_{ij}>n$, where $n$ is a predefined threshold. A muon is assigned to a momentum class according to the fraction of voxels in which its behavior is anomalous. 
 
To define the proper threshold $n$ and to check the effectiveness of the algorithm in selecting low energy muons we used simulated data generated with the GEANT4 package~\cite{agostinelli}. 
Muon detectors, support structure and lead and iron blocks are carefully described in the simulation setup.
The scrap metal is simulated with a collection of small iron cubes with mass density randomly chosen between 0 and the iron density, with an average value of $\sim 10^3$ kg/m$^3$, corresponding to the measured density of the iron scraps in our container. The detector behavior is accurately described, including the experimental error in the drift time measurements. The momentum of the simulated muons is generated according to the experimental measurements of cosmic rays angular and momentum spectra~\cite{cosmics}. 
\begin{figure}[htbp!]
\centering
\includegraphics[width=0.6\linewidth]{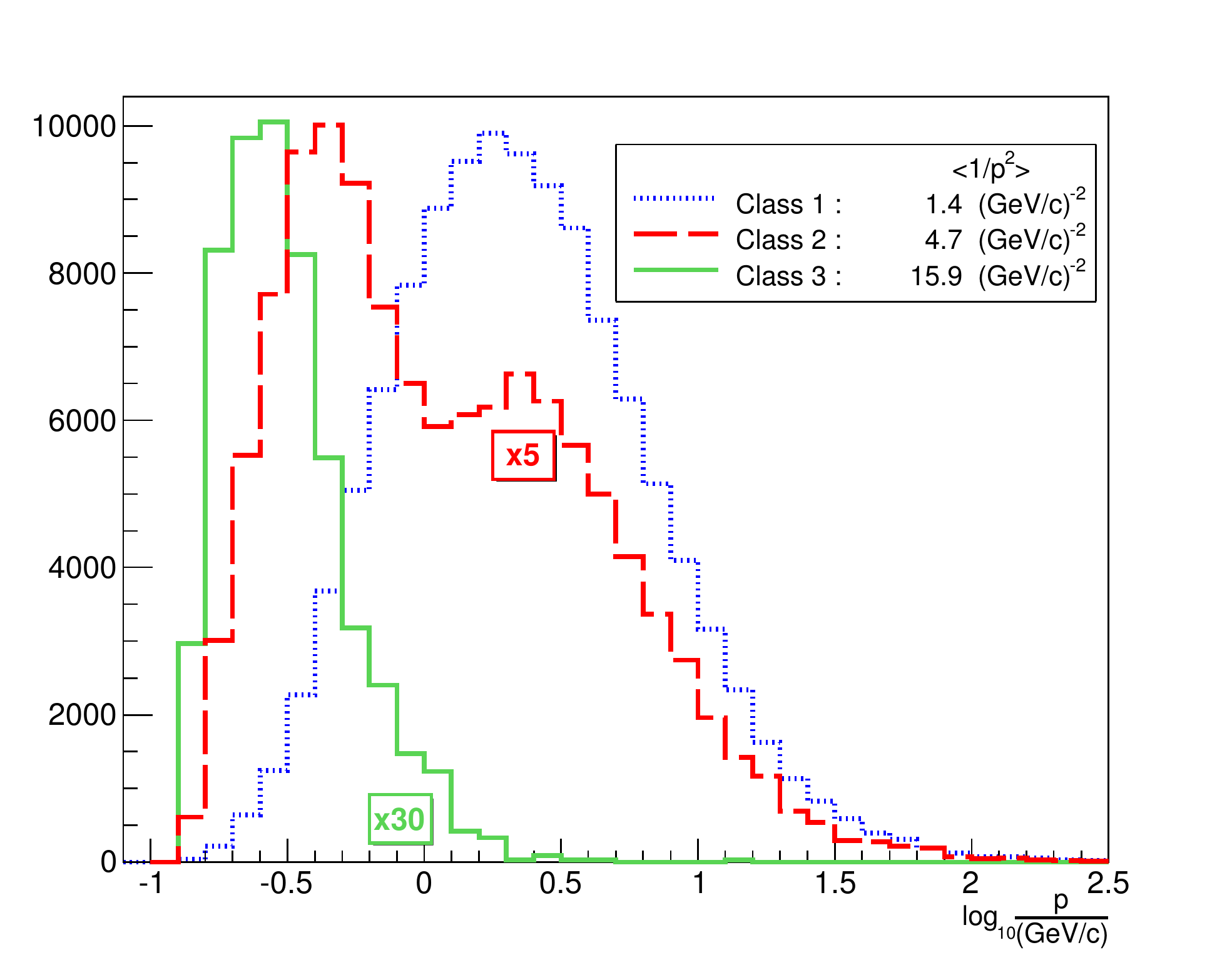}
\caption[Momentum distributions for $\Sij$ classes]
{Muon momentum distributions for the three momentum classes described in the text. Class 1: dotted line, Class 2: dashed line, Class 3: solid line. Plots entries are rescaled according to the labels. The values of $\left\langle 1/p^2\right\rangle$ for each class are reported.}
\label{fig:momentum_classes}
\end{figure}
We studied several classification criteria to select muons with different average momentum. The final choice was to use three different classes, defined as:

- Class 1: 	less than 30$\%$ of voxels in the track with $r_{ij}>3$

- Class 2:	more than 30$\%$ of voxels in the track with $r_{ij}>3$ but less than 30$\%$ with $r_{ij}>30$

- Class 3:	more than 30$\%$ of voxels in the track with $r_{ij}>30$\\
Fig.~\ref{fig:momentum_classes} shows the momentum distribution for the three muon classes.
\begin{figure*}[htbp!]
\centerline{
\subfloat[]{ \includegraphics[width=0.5\textwidth]{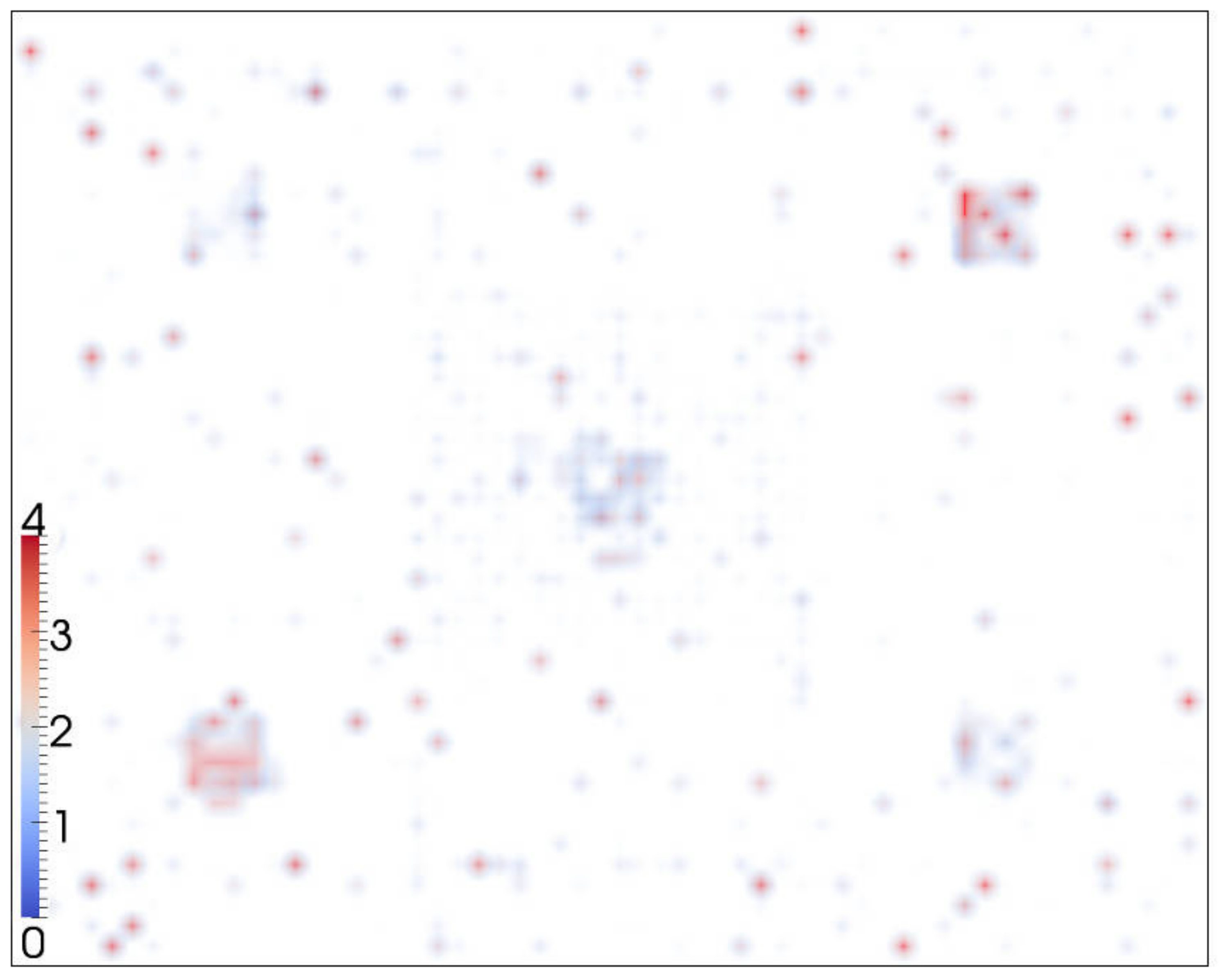} \label{fig:20m_filters_a} }
\hfill
\subfloat[]{ \includegraphics[width=0.5\textwidth]{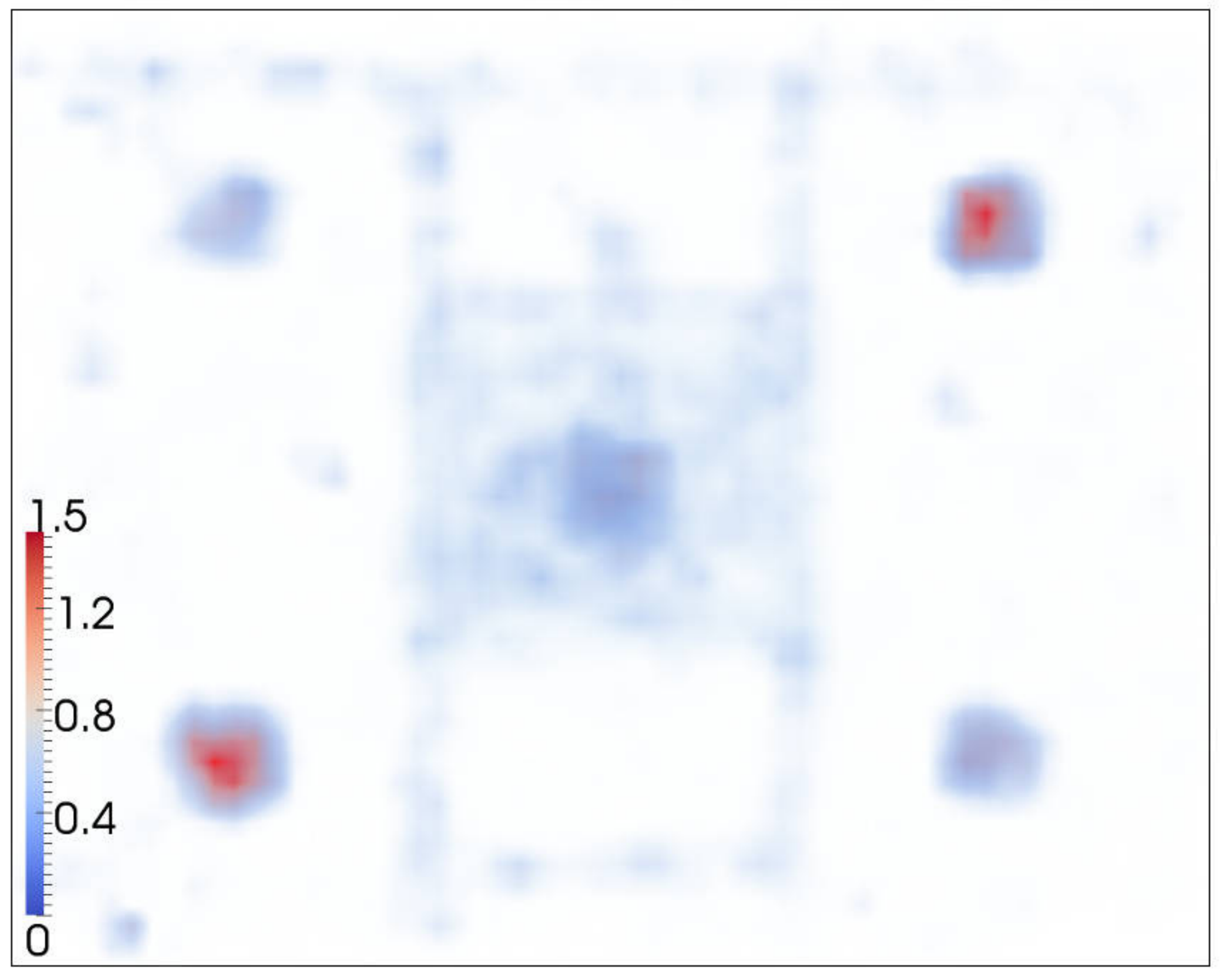} \label{fig:20m_filters_b} }
}
\centerline{
\subfloat[]{ \includegraphics[width=0.5\textwidth]{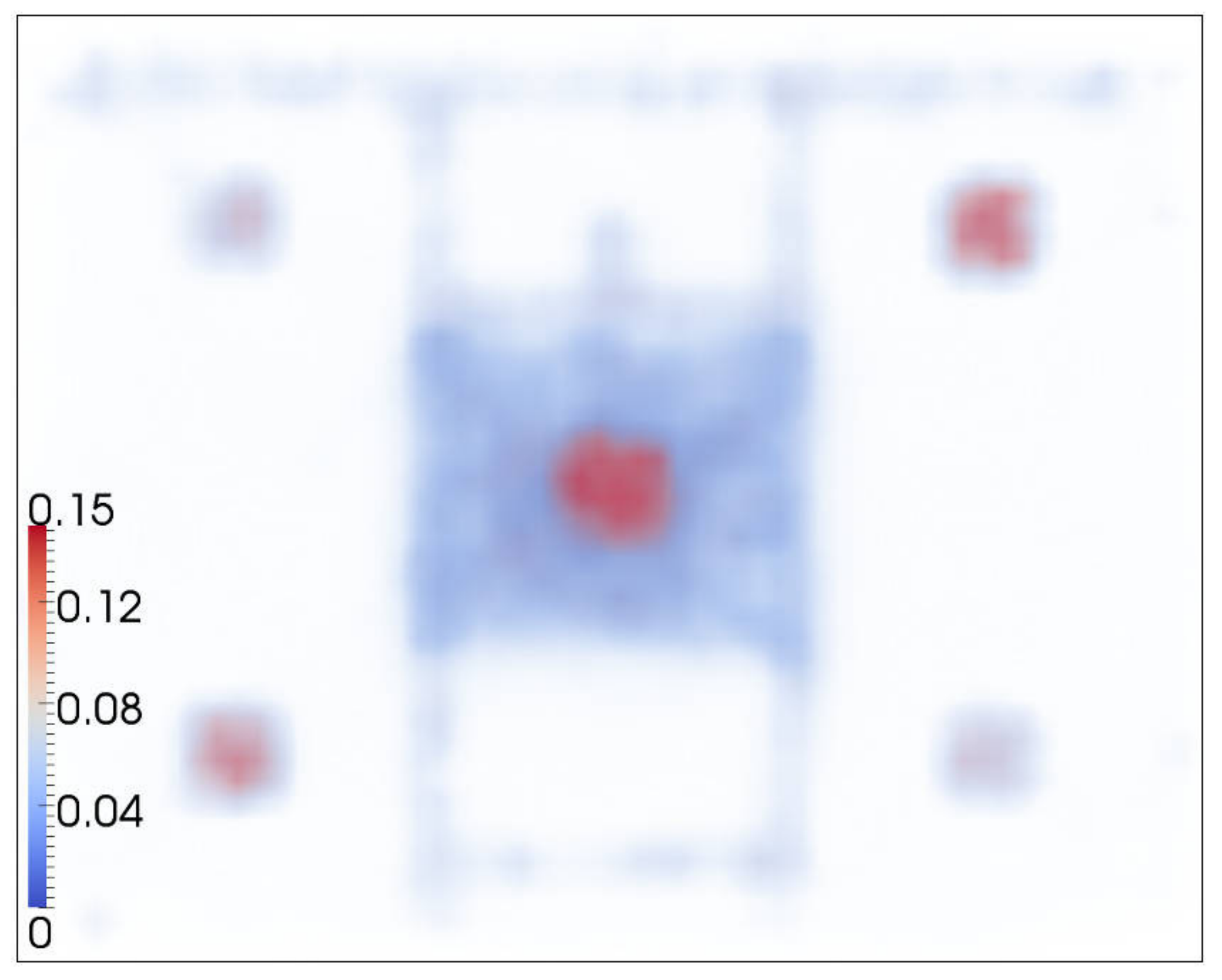} \label{fig:20m_filters_c} }
\hfill
\subfloat[]{ \includegraphics[width=0.5\textwidth]{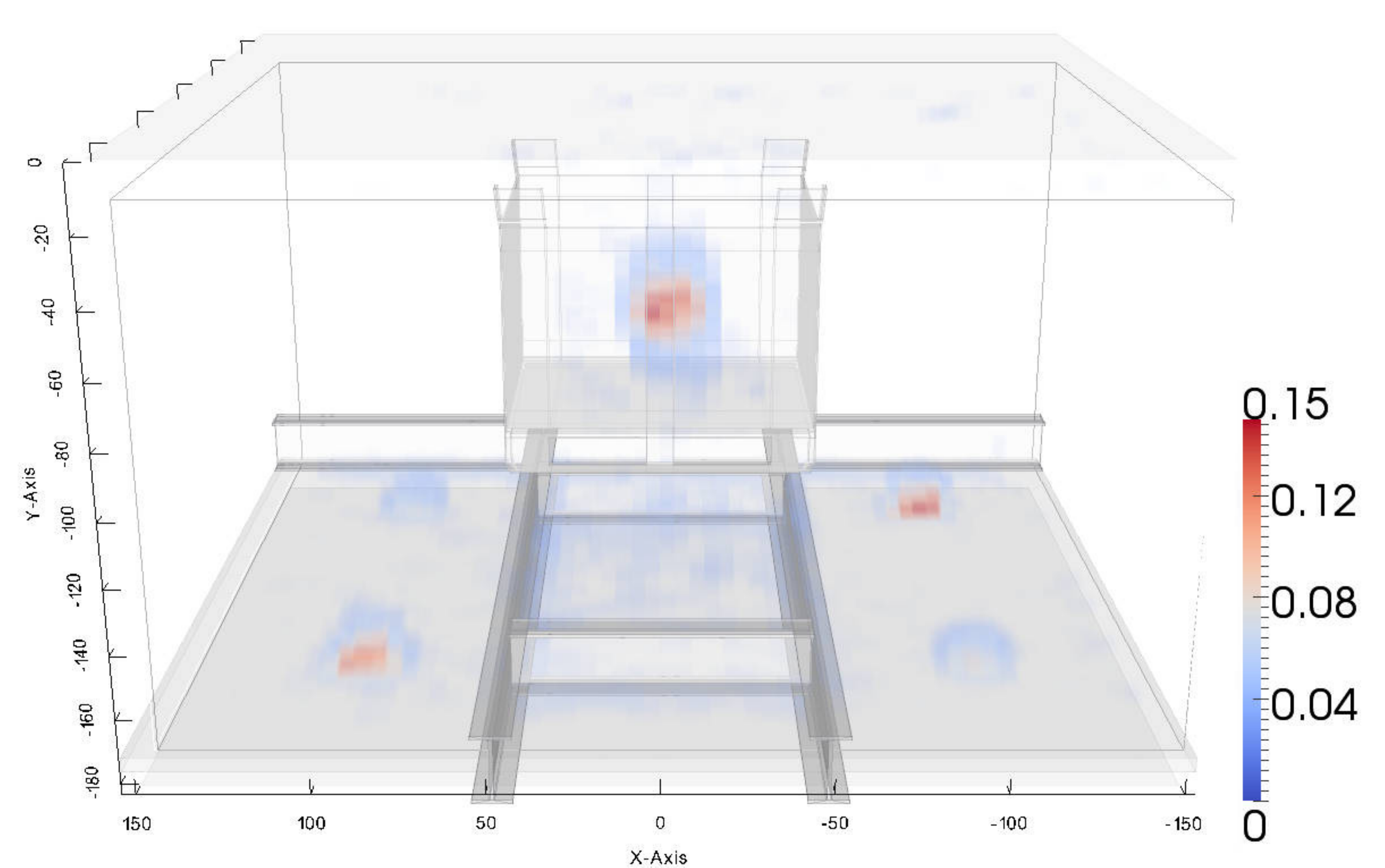} \label{fig:20m_filters_d} }
}
\caption[20 minutes reconstructions with filters]
{Tomographic reconstructions with different noise reduction techniques: (a) no filter applied, (b) $\alpha$--trim filter with a 27~voxel cubic mask and $\alpha=1/27$, (c) and (d) $\Sij$ momentum classification, as described in the text, combined with $\alpha$--trim filter. Acquisition time 7~min, voxel size 5 cm.}
\label{fig:20mfilters}
\end{figure*}
The values of $\left\langle 1/p^2\right\rangle$ 
for the three classes are: 1.4 (GeV/c)$^{-2}$, 4.7 (GeV/c)$^{-2}$, and 15.9 (GeV/c)$^{-2}$.
These values are used when the algorithm is applied to experimental data.

We show in Fig.~\ref{fig:20mfilters} the rendering of the scaled scattering density distribution of a single data sample obtained with a 7~min inspection time to visualize the effect of the different techniques described so far. The high statistics allows the use of smaller size voxels ($5$~cm).

\begin{figure*}[htbp!]
\centerline{
\subfloat[]{ \includegraphics[width=0.5\textwidth]{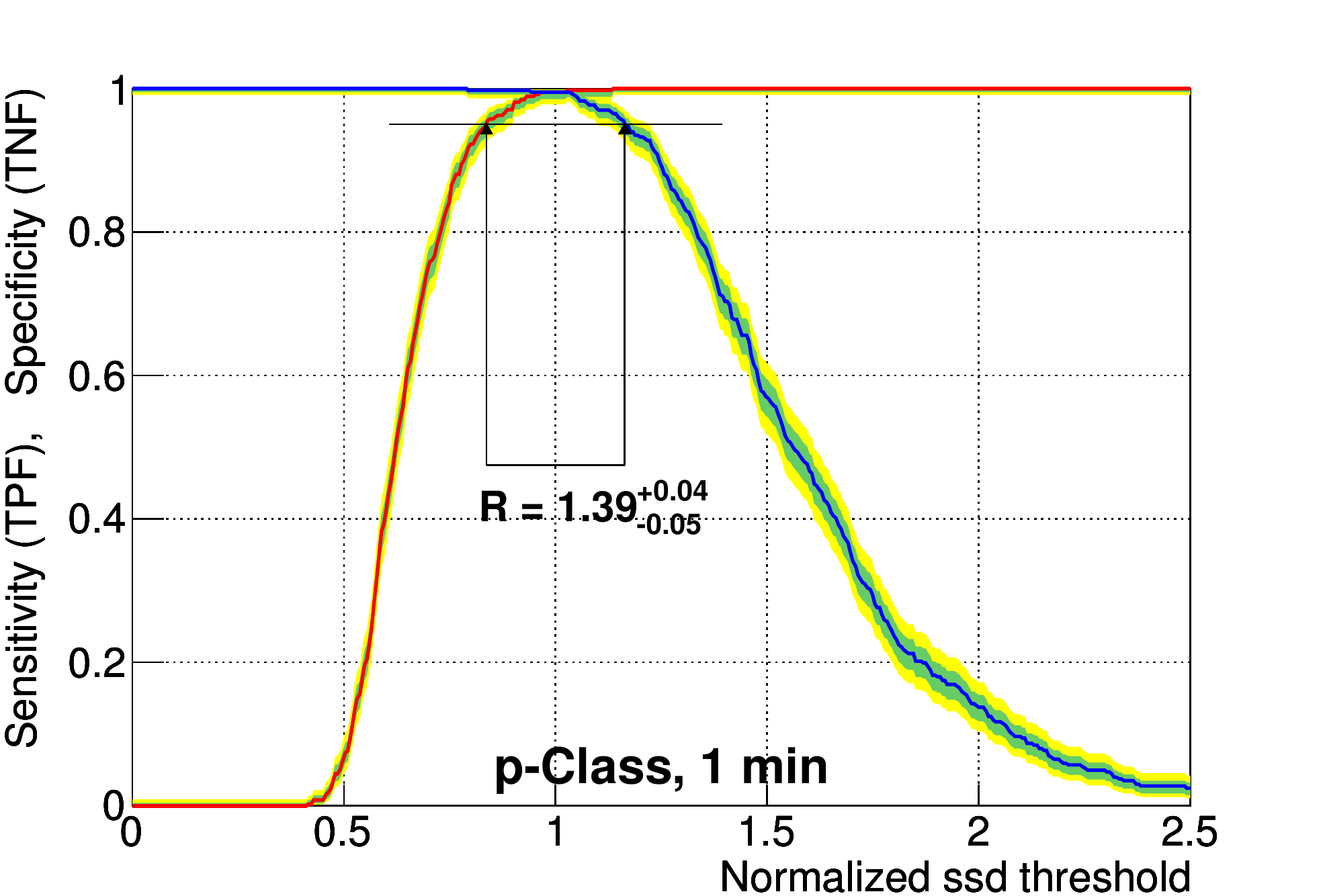} \label{fig:1minTPN_a} }
\hfill
\subfloat[]{ \includegraphics[width=0.5\textwidth]{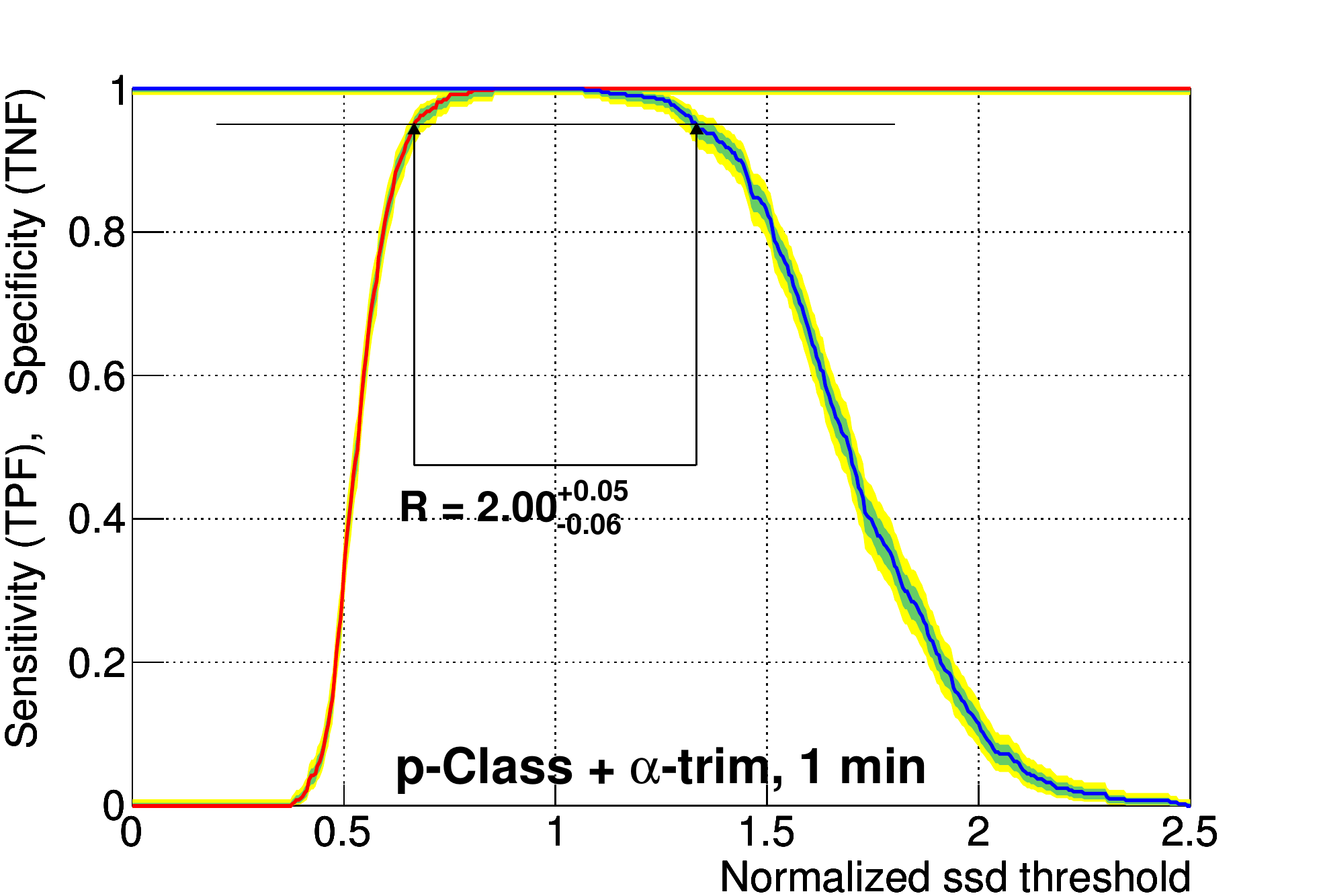} \label{fig:1minTPN_b} }
}
\caption[TPN plots 1 minute]
{(a) TPN plot using the momentum classification algorithm described in the text; (b) TPN plot adding the $\alpha$-trim filter. Inspection time 1~min.}
\label{fig:1minTPN}
\end{figure*}

The TPN plots for an inspection time of 1~min are shown in Fig.~\ref{fig:1minTPN}.
In Fig.~\ref{fig:1minTPN}~(a) the momentum classification procedure alone is applied: the separation ratio is $R=1.39^{+0.04}_{-0.05}$, slightly better than the result obtained with the default $\alpha$-trim filter shown in Fig.~\ref{fig:5-1minTPN}~(d).
Fig.~\ref{fig:1minTPN}~(b) shows TPN plots obtained applying the $\alpha$-trim filter to images reconstructed using the momentum classification procedure. The improvement is quite evident, since it leads to $R=2.00^{+0.05}_{-0.06}$.

\begin{figure*}[htbp!]
\centerline{
\subfloat[]{ \includegraphics[width=0.5\textwidth]{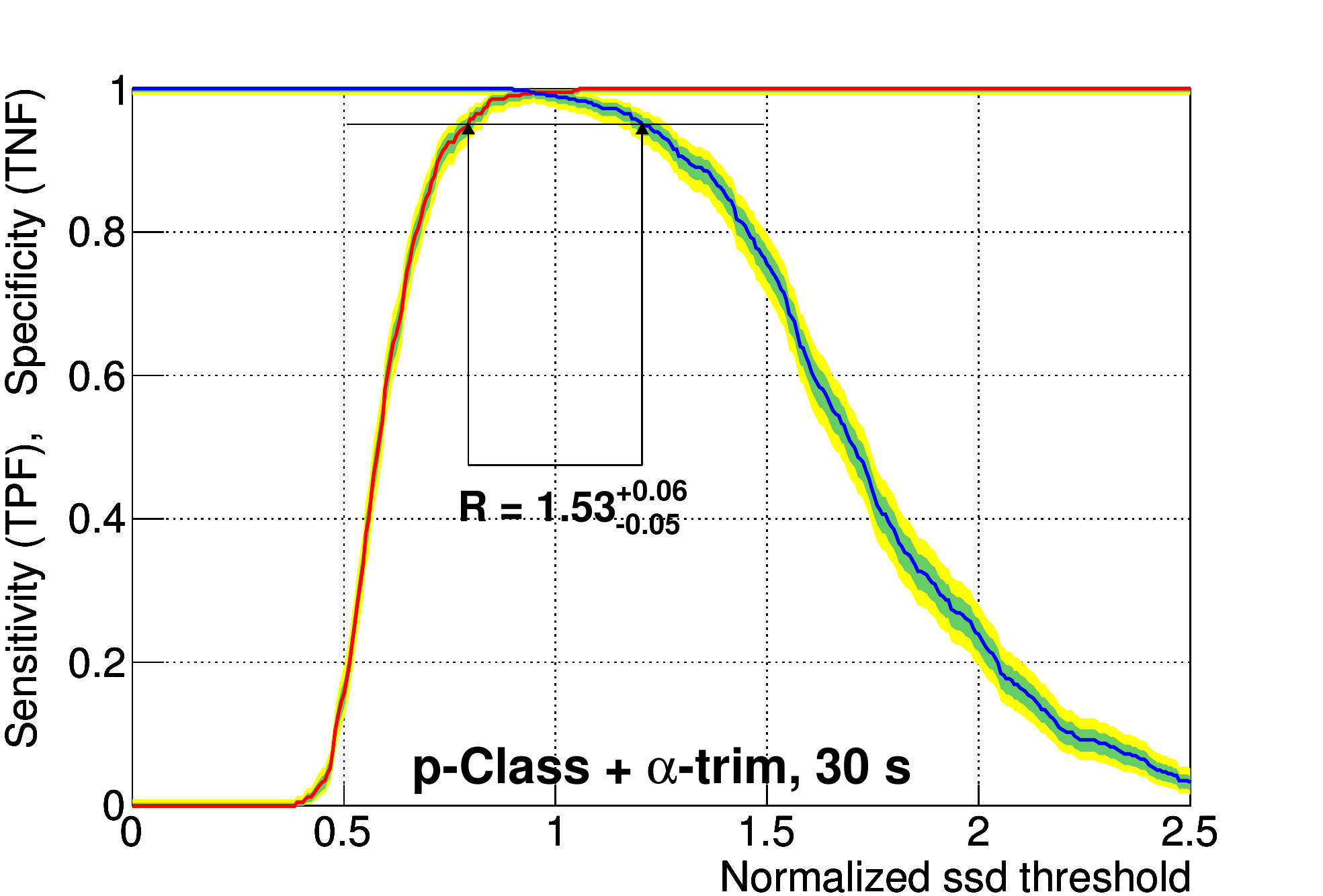} \label{fig:30-15secTPN_a} }
\hfill
\subfloat[]{ \includegraphics[width=0.5\textwidth]{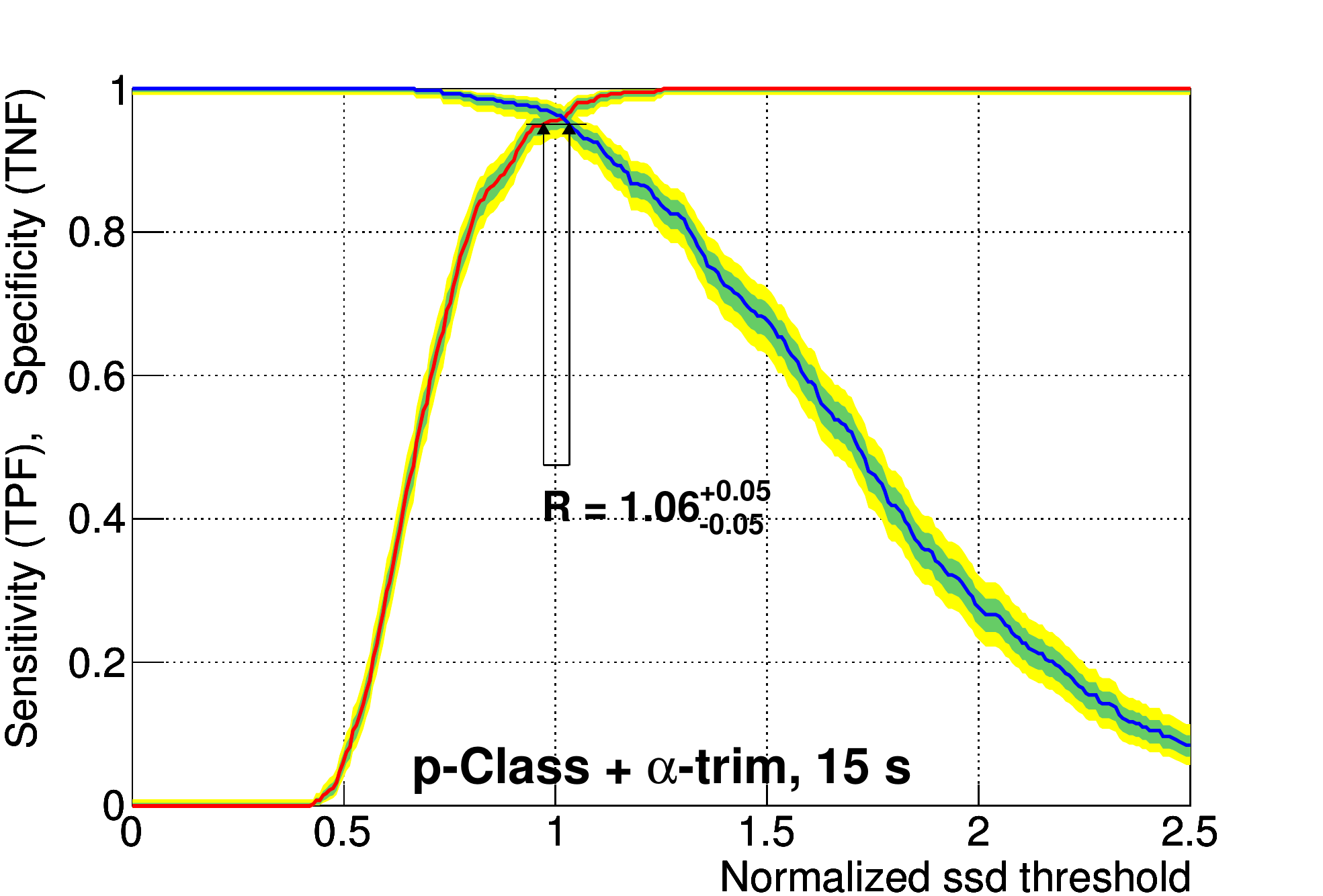} \label{fig:30-15secTPN_b} }
}
\caption[TPN plots 30-15 seconds]
{TPN plots using the procedure of Fig.~\ref{fig:1minTPN}~(b) but with shorter inspection times: (a) 30~s, (b) 15~s.}
\label{fig:30-15secTPN}
\end{figure*}

Since this separation ratio is remarkable, we analyze shorter inspection times.
An acquisition time of 30~s  still shows a useful separation ratio: $R=1.53^{+0.06}_{-0.05}$ as evident from Fig.~\ref{fig:30-15secTPN}~(a). At 15~s, Fig.~\ref{fig:30-15secTPN}~(b), the ratio drops to $R=1.06^{+0.05}_{-0.05}$.

\section{Comparison with published filters}

Scattering density fluctuations were pinpointed to be a serious issue in most published papers on muon tomography. In some cases recipes to reduce the noise have been proposed.
However, the suggested algorithms were tested only on simulated data (or in very small prototypes) in all the published analysis we are aware of, sometimes assuming perfect knowledge of the muon momentum \cite{schultz}. We analyze real data from a medium--sized prototype and we propose a comparison between those algorithms and ours.

\begin{figure*}[htbp!]
\centerline{
\subfloat[]{ \includegraphics[width=0.5\textwidth]{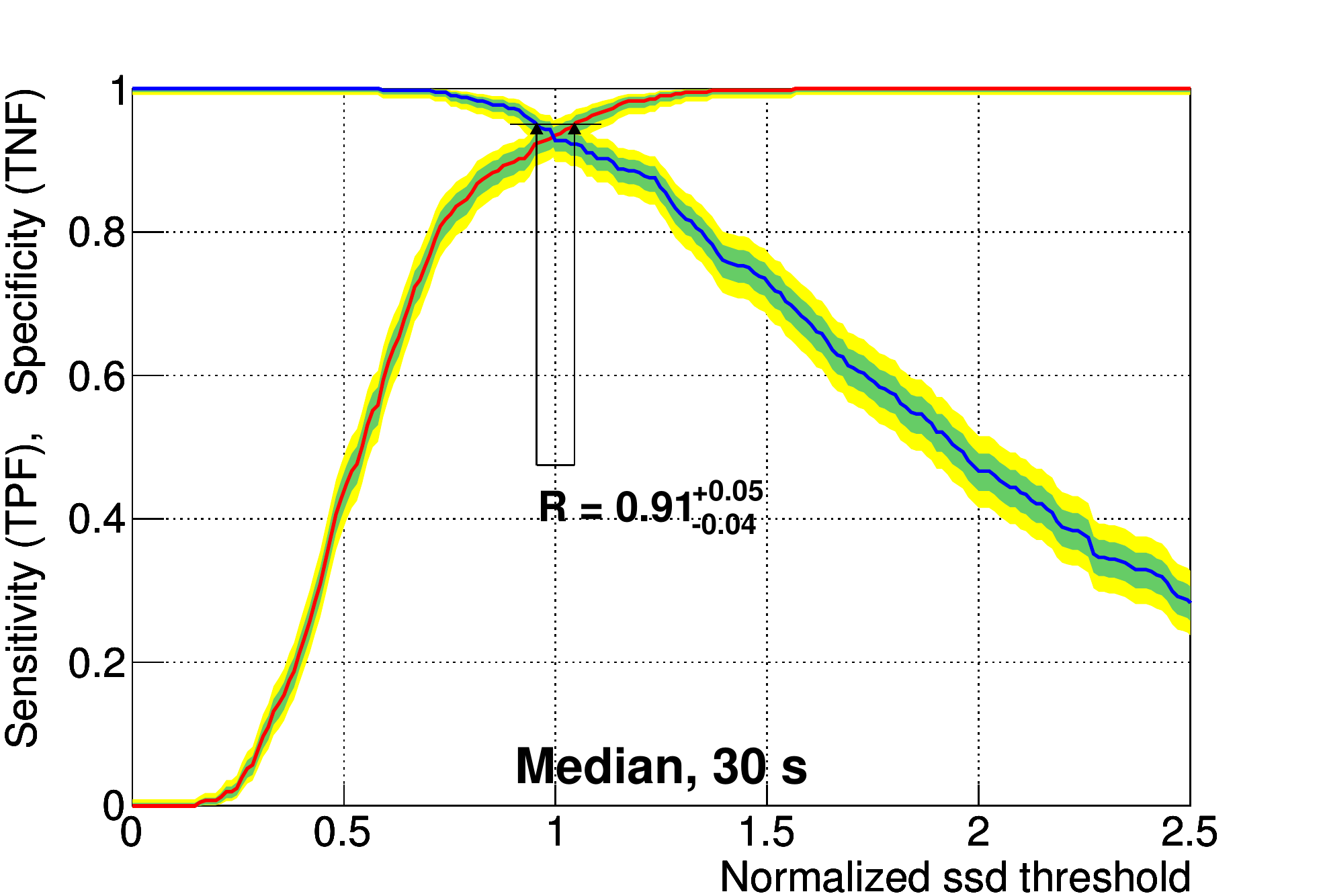} \label{fig:30secTPNmedian_a} }
\hfill
\subfloat[]{ \includegraphics[width=0.5\textwidth]{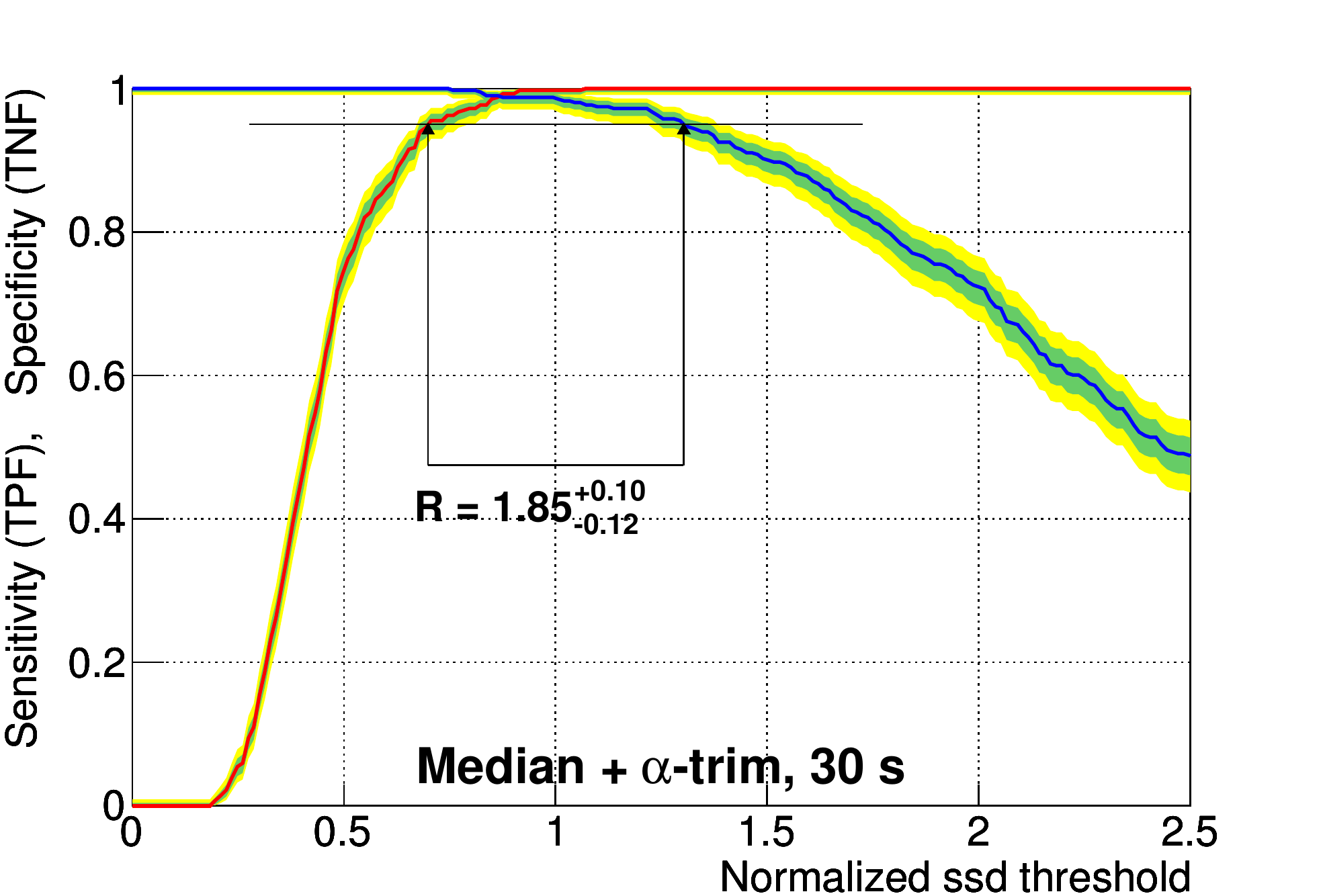} \label{fig:30secTPNmedian_b} }
}
\caption[TPN plots with $\Sij$ median 30~s]
{(a) TPN plot using the median of $\Sij$ in the image reconstruction process;  (b) same as (a) adding the $\alpha$-trim filter. Inspection time for both figures is $30$~s.}
\label{fig:30secTPNmedian}
\end{figure*}

The replacement of the average of $\Sij$ with its median value during the iterative process of the likelihood maximization, is proposed in Ref.~\cite{schultz}.
Equation~\ref{eq:lambda} is replaced with
\begin{equation}
\label{eq:lambda_median}
\lambda _j^{\left( {n + 1} \right)} = \lambda _j^{\left( n \right)} + {\left( {\lambda _j^{\left( n \right)}} \right)^2}
\mbox{median}_{\mbox{~}\muL_{ij} \ne 0}\left( {\Sij} \right).
\end{equation}
Fig.~\ref{fig:30secTPNmedian}~(a) shows that the result of the median algorithm is considerably worse than our best result, shown in Fig.\ref{fig:30-15secTPN}~(a). However, applying the post-processing $\alpha$--trim filter to the images improves the TPN plot, as shown in Fig.~\ref{fig:30secTPNmedian}~(b). 
The separation ratio, $R=1.85^{+0.10}_{-0.12}$, is now marginally better than the one in  Fig.\ref{fig:30-15secTPN}~(a).

Anyway, it is worth noticing that the median algorithm is very demanding both in computing time and memory consumption.

\begin{figure}[htbp!]
\centerline{
\subfloat[]{ \includegraphics[width=0.5\textwidth]{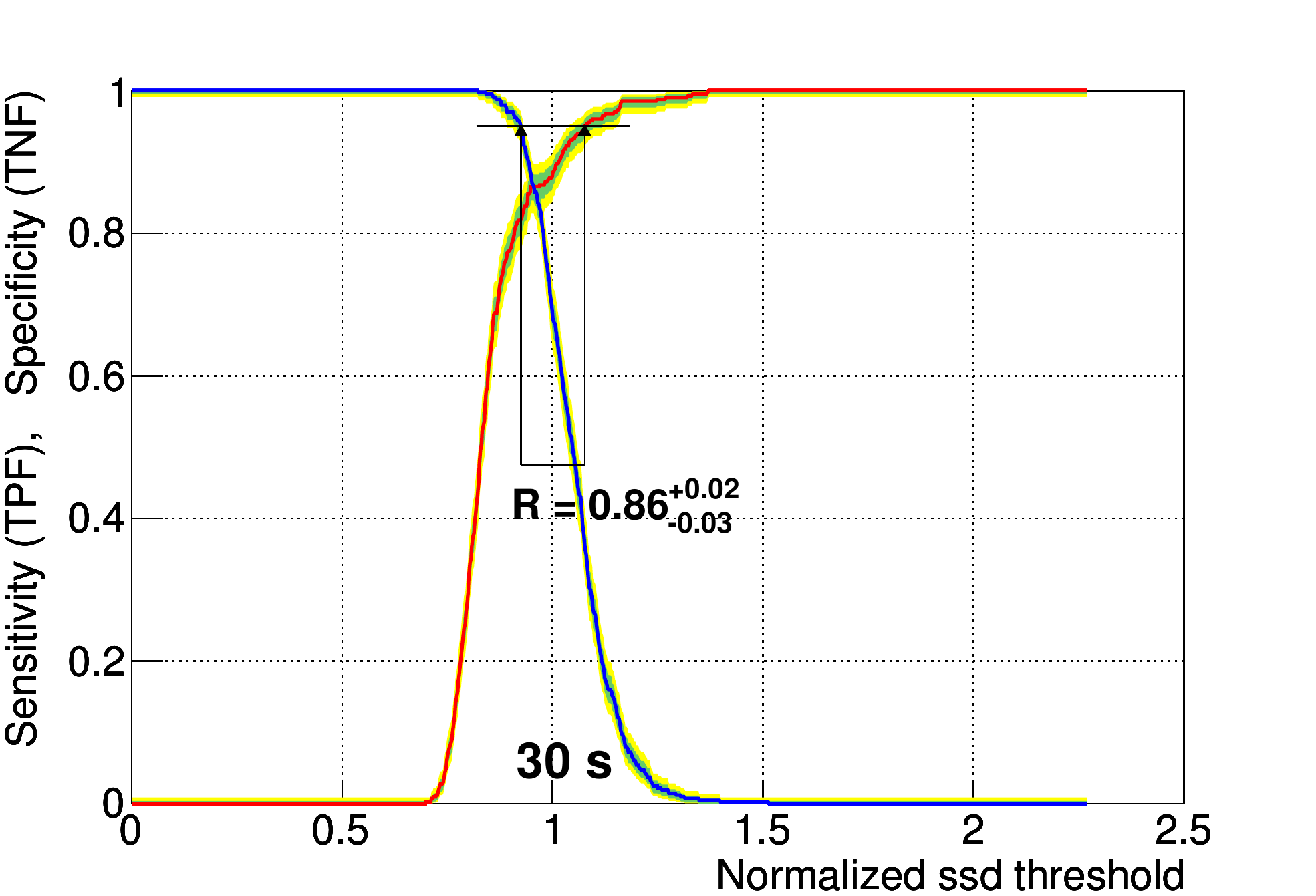} \label{fig:priorTPN_b} }
\hfill
\subfloat[]{ \includegraphics[width=0.5\textwidth]{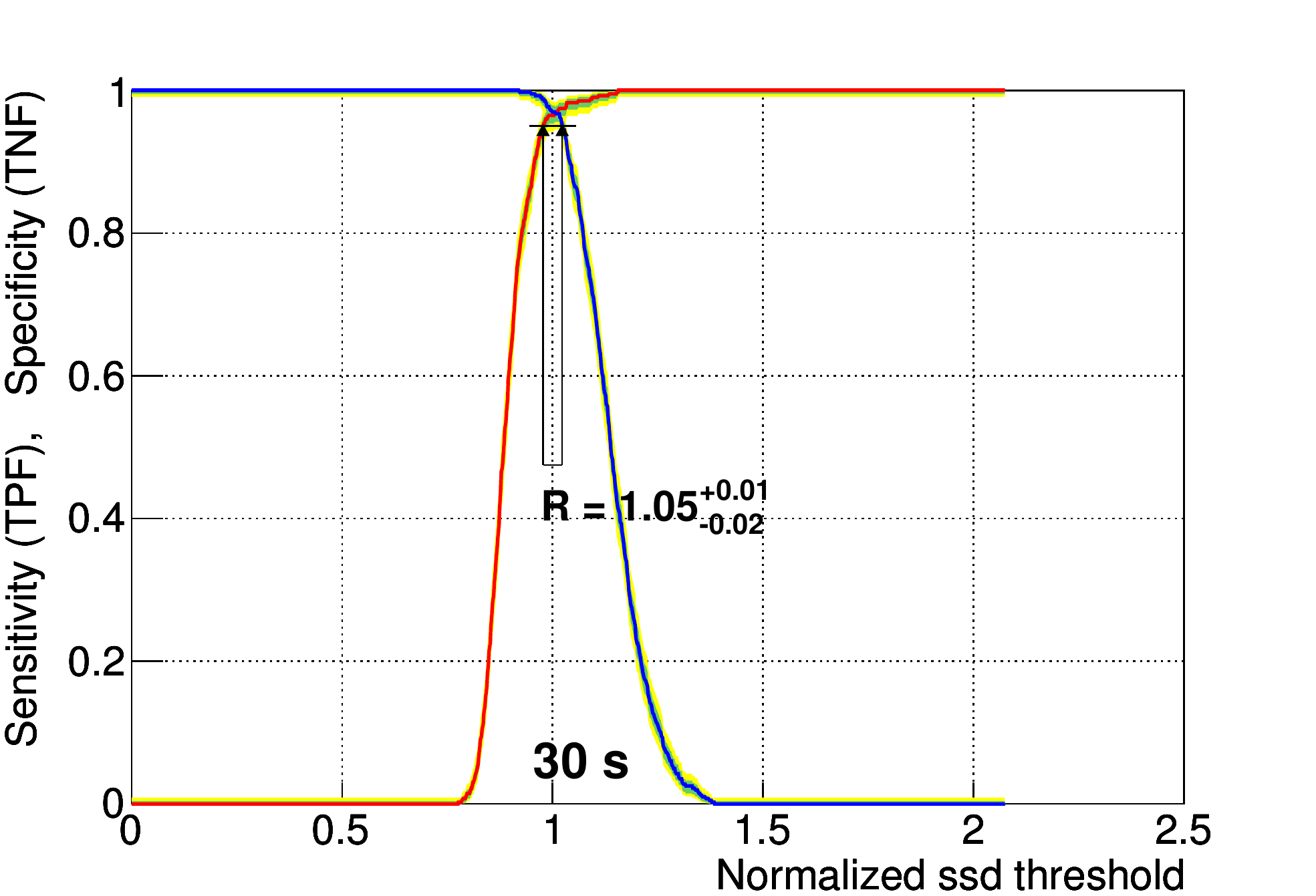} \label{fig:priorTPN_a} }
}
\caption[TPN plots with prior]
{ TPN plots using Bayesian Gaussian prior with $\epsilon=4\cdot10^{-4}$~rad$^{2}$/cm; reconstructed images smoothed with (a)~Gaussian  filter, (b) $\alpha$--trim filter. Inspection time for both figures is $30$~s.}
\label{fig:priorTPN}
\end{figure}

Ref.~\cite{wang} proposes a modification of the likelihood function. The scattering densities are bound to be distributed according to a Bayesian prior probability function. Best results are obtained with a Gaussian prior: 
\begin{equation}
p(\lambda_j)=A \cdot e^{-\lambda_j^{2}/2\epsilon^{2}}
\end{equation}
where A is a normalization factor and $\epsilon$ a free parameter. 
The scattering density values of different voxels are assumed to be independent. 
In addition ref.~\cite{wang} shows that smoothing the reconstructed images with a Gaussian filter improves the ROC curves.
When this recipe is applied to our data, a value of $\epsilon=4\cdot 10^{-4}$ ~rad$^{2}$/cm gives the best result\footnote{This value of $\epsilon$ corresponds to $\beta = 10^{4} (\mbox{mrad}^2/\mbox{cm})^{-2}$ using the formalism of Ref.~\cite{wang}}. The TPN plot is shown in Fig.~\ref{fig:priorTPN}~(a). Replacing the Gaussian filter with the $\alpha$--trim filter the TPN plot of Fig.~\ref{fig:priorTPN}~(b) is obtained. The separation ratio increases from $R=0.86^{+0.02}_{-0.03}$ to $R=1.05^{+0.01}_{-0.02}$, a value still worse then the one obtained with momentum classification method. In this case the two curves are very steep and consequently the errors on R are small. However the choice of the operational threshold becomes critical.

It is worth mentioning that both Ref.~\cite{schultz} and~\cite{wang} are dealing with simulated data only. Our results are drawn from experimental data, and do not assume the exact knowledge of the muon momenta.


\section{Conclusions}

In this paper an innovative technique is proposed to handle the problem of the statistical noise in muon tomography images. This approach greatly improves the performance of the technique in detecting dense materials in extended volumes.

The method consists both in a technique to classify muon momenta from the scattering data themselves and in a post--processing filtering of the reconstructed images.
The method has been applied to experimental data collected using the muon tomography prototype installed in the INFN LNL. 

In the examined configuration a lead block of 25$\times$25$\times$20 cm$^3$ was hidden in a container filled with iron scraps. 
The proposed technique is able to detect the lead block in 30~s, with 
sensitivity and specificity both close to  
$1$. With this performance the system is suitable for commercial application.


\end{document}